\documentclass[aps,prl,floatfix,twocolumn,superscriptaddress,showpacs,amsmath,amssymb,showpacs,nolongbibliography,footinbib]{revtex4-2}

\usepackage{graphicx}
\usepackage{epstopdf}
\usepackage{epsfig}
\usepackage{epsf}
\usepackage{url}
\usepackage[USenglish]{babel}
\usepackage{hyperref}
\def\bcen{\begin{center}}
\def\ecen{\end{center}}
\allowdisplaybreaks
\renewcommand\[{\begin{equation}}
\renewcommand\]{\end{equation}}
\usepackage{verbatim}
\usepackage{xcolor}
\usepackage{natbib}
\usepackage{amsmath, nccmath}
\usepackage{dcolumn}
\usepackage{bm}
\usepackage{bbm}
\usepackage{lipsum}
\usepackage[normalem]{ulem}  

\usepackage{orcidlink}

\begin{document}

\title{Substrate and cation engineering for optimizing superconductivity \\ in infinite-layer nickelates}
\author{Viktor Christiansson\orcidlink{0000-0002-9002-0827}}
\affiliation{Institute of Solid State Physics, TU Wien, 1040 Vienna, Austria}

\author{Karsten Held\orcidlink{0000-0001-5984-8549}}
\affiliation{Institute of Solid State Physics, TU Wien, 1040 Vienna, Austria}

\begin{abstract}
In a  recent experiment [Nature {\bf 642}, 58 (2025)], a new record for the superconducting critical temperature $T_c$ among infinite-layer nickelates has been reported in doped SmNiO$_2$. 
Here, we use the cutting-edge dynamical vertex approximation (D$\Gamma$A), and qualitatively as well as quantitatively reproduce the $T_c$~vs.~doping dome for this compound.
Encouraged by this, we go further and identify a path towards realizing even higher $T_c$'s by changing the cation along the line Nd$\rightarrow$Sm$\rightarrow$Y$\rightarrow$Lu with matching substrates. The successively smaller cation radius allows for smaller lattice constants of the substrate. This in turn
increases the in-plane hopping and thus  eventually $T_c$.
\end{abstract}
\date{\today}

\maketitle

\begin{figure}[t]
	\centering
\includegraphics[angle=0, width=1 \columnwidth]{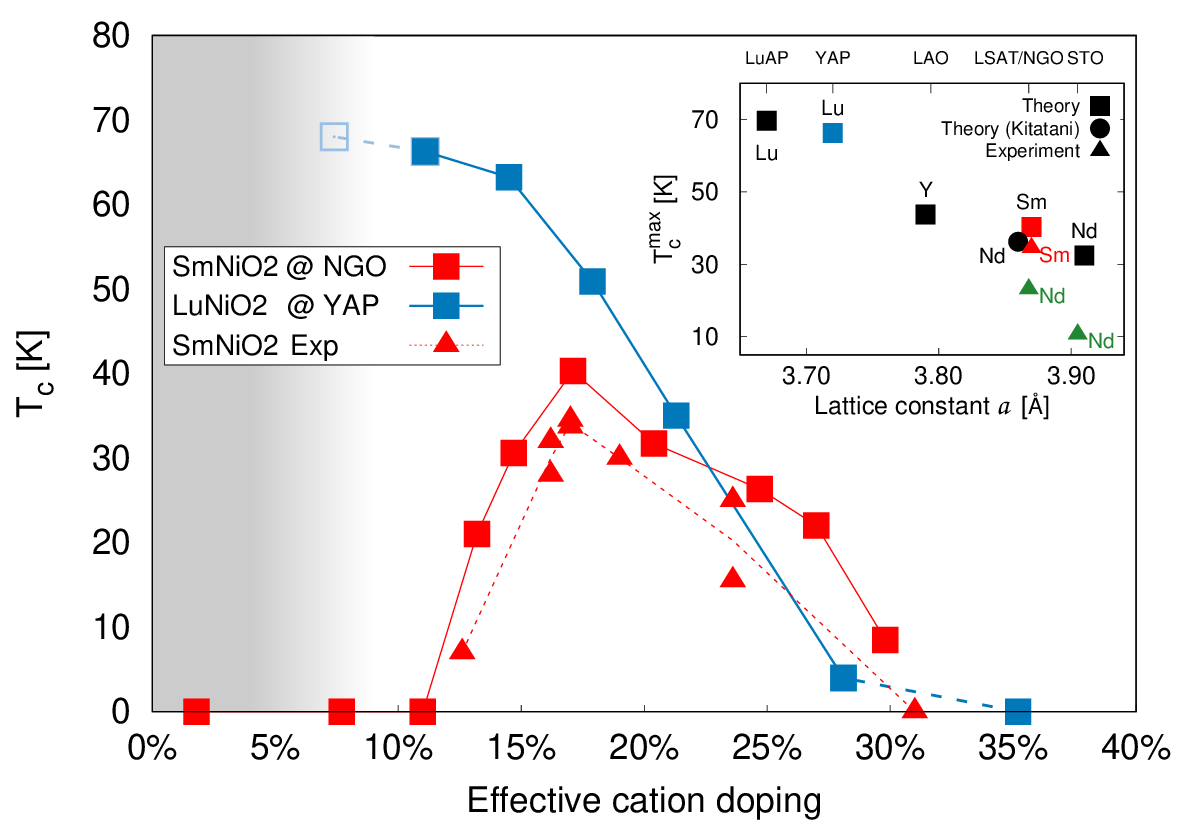}
\caption{\label{Fig:Tc_doping} 
  $T_c$ vs. effective doping of SmNiO$_2$ (experiment: red triangles \cite{Chow2025}; and theory: red squares) and LuNiO$_2$ (conjectured: open blue squares).
  The shaded area indicates a region where a competing antiferromagentic order could be expected, see text. Inset: Maximal $T_c$ vs.\ in-plane lattice constant for various substrates and matching
  cations $R$ as listed in Table~\ref{Table:hoppings}; calculated values are shown with squares (red and blue refer to the compounds shown in the main figure), and
  the experimental data is taken from \cite{Li2020} for NdNiO$_2$ on STO and \cite{Lee2023} for NdNiO$_2$ on LSAT (green triangles), and ~\cite{Chow2025} for SmNiO$_2$ on NGO (red triangle). The theoretical $T_C$ marked by a black circle is from \cite{Kitatani2020}.
  }
\end{figure}

{\it Introduction.}
With the discovery of superconductivity in the latest member of the infinite-layer (IL) nickelate family,  SmNiO$_2$, a new record was set for IL nickelates: a critical temperature $T_c$ close to 40 K \cite{Chow2025} at ambient conditions (see Fig.~\ref{Fig:Tc_doping}).
While IL nickelates had been conjectured  to be an analog of cuprates \cite{Anisimov1999} 25 years ago, it was the more recent discovery of superconductivity in Sr-doped NdNiO$_2$ with $T_c\sim10$ K \cite{li2019superconductivity} that set the stage for an enormous experimental and theoretical effort.
Since then, several members of the IL family (with general formula $R$NiO$_2$) have been synthesized and found to exhibit superconductivity:
$R=$ Pr ($T_c\sim14$ K \cite{Osada2020a,Osada2020b}), La ($T_c\sim9$ K \cite{Osada2021,Zeng2022}). A higher critical temperature has been observed when applying pressure to PrNiO$_2$ ($T_c\sim31$ K  \cite{Wang2022}). 
Taking a  broader perspective, this effort has also led to the discovery of superconductivity  in a range of further nickelates, including the finite-layer-$n$ cousins of IL nickelates $R$$_{n+1}$Ni$_n$O$_{2n+2}$ \cite{pan2021} and Ruddlesden-Popper (RP) type  nickelates $R$$_{n+1}$Ni$_n$O$_{3n+1}$ \cite{Sun2023}.

 Whereas the RP nickelates with their 3$d^{7+1/n}$ electronic configuration clearly require a multiorbital description, the relevant orbitals of the low-energy 
 physics of the IL phase with a doped 3$d^9$ configuration have been hotly debated early on. Some proponents argued for the multiorbital nature \cite{Lechermann2020a,Petocchi2020,Lechermann2020b,Christiansson2023}, while others proposed that a minimal single-band model plus decoupled pockets was sufficient \cite{Kitatani2020,Karp2020,Pascut2023,Held2022}. Recent experimental evidence is in favor of the latter view with ARPES showing only the Ni 3$d_{x^2-y^2}$ orbital plus an $A$ pocket \cite{Sun2024,Ding2024}.
Noteworthy, using such an effective single-band Ni $3d_{x^2-y^2}$ model, dynamical vertex approximation (D$\Gamma$A) \cite{Toschi2007,RMPVertex} calculations~\cite{Kitatani2020} were able to accurately predict
the $T_c$~vs.~doping phase diagram, later confirmed in experiment after high quality films had been synthesized \cite{Lee2023}. Also angular resolved photoemission spectroscopy and resonant inelastic x-ray scattering (RIXS) were found to agree well with the D$\Gamma$A calculations \cite{Lu2021,Worm2024,Sun2024,Ding2024,Si2024}. These describe the (quasi)particles and their
spin fluctuation pairing glue, respectively, which enter in the  D$\Gamma$A calculation of $T_c$.
Taken altogether, this gives us quite some confidence that an accurate theoretical modeling of IL nickelates has been achieved.

The recent discovery of a record high $T_c$ in doped SmNiO$_2$ \cite{Chow2025} has once again highlighted the importance of the IL nickelate family.
One possibility for the higher $T_c$ is that the grown films might be simply of better quality. Indeed, due to their complicated synthesis, IL nickelates are plagued by defects; and synthesizing better films led to a substantial increase in $T_c$ for NdNiO$_2$\cite{Lee2023}, and consequently better agreement with theory~\cite{Kitatani2020}.
However, the residual resistivity of the SmNiO$_2$ films reported in Ref.~\cite{Chow2025}  ($T_c\gtrsim35$ K) is similar to the most defect-free NdNiO$_2$ films ($T_c\sim23$  K \cite{Lee2023}). This indicates that it is likely not the full story.

In this Letter, we study a range of rare-earth $R$ infinite-layer nickelates, some of which are already experimentally realized ($R=$ Nd and Sm), and others still need to be synthesized ($R=$Y and Lu). 
To more directly connect to experiment, we consider common substrates with lattice constants that match within a few percent to $R$NiO$_2$ and $R$NiO$_3$.
We downfold the electronic structure to an effective one-band Ni-$3d_{x^2-y^2}$ model with an additional electron reservoir to account for the $R$ self-doping. Our main finding is that smaller cation $R$ radii allow for substantially larger $T_c$'s, and that the recently discovered SmNiO$_2$ with $T_c$ close to 40\,K is only the first step on this route to higher $T_c$'s in IL nickelates, as summarized in Fig.~\ref{Fig:Tc_doping}.

{\it Method}
Experimentally superconducting IL nickelates have so far always been grown as thin films on various substrates. More specifically, first  $R$NiO$_3$ films are grown and these
are subsequently reduced to 
$R$NiO$_2$ \cite{Lee2020}.
To conform with this experimental recipe, we first calculate the relaxed lattice parameters in the bulk IL phase (space group $P4/mmm$) by density functional theory (DFT) using the generalized gradient approximation (GGA) with Perdew-Burke-Ernzerhof (PBE) exchange correlation functional \cite{Perdew96},
as implemented in the \textsc{WIEN2k} code \cite{blaha2001wien2k}. For computational details, see the Supplemental material (SM)~\cite{SM}.
We then compare these bulk lattice parameters to potential substrates that match within  $\lesssim3$\% difference in the in-plane lattice parameters $a=b$. Finally, we also compare 
the (slightly smaller) in-plane lattice parameters of the perovskite starting point that need to match with the substrate lattice parameters within a few percent as well, 
for the initial epitaxial growth to be experimentally viable.

As we are here interested in thick films and not in interface and surface effects that affect a very few layers close to the interface or surface (for these see, e.g.,~\cite{Ortiz2021b,Verhoff2025}), we do a ``bulk-like'' calculation
with the in-plane lattice constant epitaxilly fixed to the $a$ lattice constant of the substrate.
For the fixed in-plane lattice constant of the substrates, we then fully relax the out-of-plane $c$-lattice constants, and we have checked that varying $c$ does not have a large effect on the resulting in-plane hopping parameters, and as such on our final results (see SM~\cite{SM}). 
The substrates and cations $R$ considered are listed together with the substrate lattice parameters $a=b$ 
in Table~\ref{Table:hoppings}.

Next, we first define a 10-orbital model by projecting onto maximally localized Wannier functions on the Ni and $R$ $d$ states employing
\textsc{wien2wannier} \cite{Kunes2010a} and \textsc{Wannier90} \cite{Pizzi2020}. We supplement this DFT-derived Wannier Hamiltonian with the same site-dependent Coulomb interactions as previously used \cite{Kitatani2020}, (i.e., a Kanamori interaction with $U_{\mathrm{Ni}}$=4.4 eV, $J_{\mathrm{Ni}}$=0.65, $U_R$=2.5 eV, and $J_R=$0.25), assuming that this is a local property and should not change significantly over the series for different $R$ atoms. We then do
a DMFT calculation at inverse temperature $\beta=40$ eV$^{-1}$ ($T=290$ K) using \textsc{w2dynamics} \cite{w2dynamics2018}. 
As will be shown below, this multi-band DMFT calculation validates that a single-orbital description with an additional electron reservoir
\footnote{This ``electron reservoir'' corresponds to the $R~5d_{xy}$ orbital and mainly acts to account for the self-doping due to the $R$ atom, see~\cite{Kitatani2020}.}
as proposed in Ref.~\cite{Kitatani2020} is an appropriate low-energy description for these various new IL nickelates. 
Thus, we subsequently define a new single-band model by projecting instead on the Ni $3d_{x^2-y^2}$ state, and calculate the superconducting  $T_c$ 
within single-band dynamical vertex approximation (D$\Gamma$A) 
using the \textsc{DGApy} code \cite{worm2023}.

\begin{figure}[t]
	\centering
    \includegraphics[angle=0, width=1.0 \columnwidth]{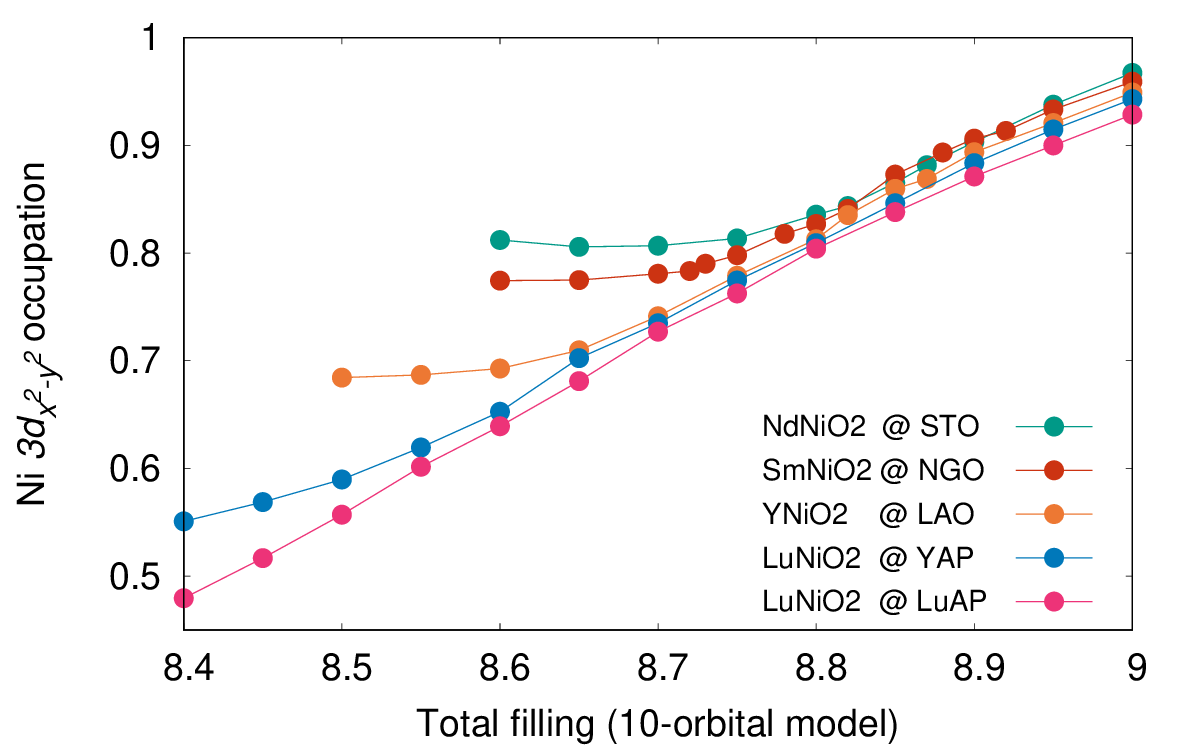}
    \caption{\label{Fig:mainFilling} Filling of the $d_{x^2-y^2}$ orbital as a function of the total filling in the 10-orbital model, see the text, at inverse temperature $\beta=40$ eV$^{-1}$ ($T=290$ K) for the different IL nickelates studied.}
\end{figure}

\begin{table*}[t]
\setlength{\tabcolsep}{4.5pt} 
\renewcommand{\arraystretch}{1.4} 
    \centering
    \caption{Considered IL nickelates and substrates. In plane lattice parameter $a$(=$b$) of the substrate,
    substrate strain for bulk $R$NiO$_2$ and  $R$NiO$_3$ \cite{pseudocubic}, as well as the single band hopping parameters and $U/t$ ratio with fixed $U=3.2$ eV.} 
    \label{Table:hoppings}
 \begin{tabular}{|ll||c  c  c|| c c  c c|}
\hline
Nickelate  @ substrate&& $a$ & strain $R$NiO$_2$ & strain $R$NiO$_3$ & $t$ [meV] & $t'$ [meV] & $t''$ [meV] & $\left|U/t\right|$  \\
\hline
\hline
NdNiO$_2$ @ SrTiO$_3$& (STO)      & 3.91 & -0.3$\%$ & -2.7$\%$  & -379 & 97 & -45 & 8.4 \\
SmNiO$_2$ @ NdGaO$_3$& (NGO)      & 3.87 & 0.2$\%$ & -1.7$\%$ & -393  & 98 & -46 & 8.1 \\
YNiO$_2$ @ LaAlO$_3$& (LAO)     & 3.79 & 1.6$\%$ & -1.0$\%$ & -419 & 94 & -46 & 7.6 \\
LuNiO$_2$ @ YAlO$_3$& (YAP)     & 3.72 & 0.2$\%$ & 2.2$\%$ & -449 & 108 & -55 & 7.1 \\
LuNiO$_2$ @ LuAlO$_3$& (LuAP)     & 3.67 & 3.4$\%$ & 1.3$\%$ & -467 & 105 & -56 & 6.9 \\
\hline
\end{tabular}
\end{table*}

\textit{DMFT results.}
Let us start here by discussing the orbital filling of the multi-orbital (DFT+)DMFT calculation. For the DFT bandstructure and the Wannier projection, as well as the 10-orbital model DMFT self-energy and spectral functions, see the SM \cite{SM}.
In the undoped parent compound the total filling of the Ni and $R$ $d$ manifold is 9 electrons. 
We find that the Ni $3d_{x^2-y^2}$ is close to half-filling, whereas the other Ni orbitals are almost completely filled, as has been previously discussed for the IL nickelates based on, e.g., DFT+DMFT \cite{Lechermann2020b,Karp2020,Kitatani2020} and $GW$+EDMFT \cite{Petocchi2020,Christiansson2023}.
To mimic the effective doping of the system without the need to consider a supercell, the typical approach is to successively reduce the filling of the model space and solve the impurity problem in DFT+DMFT self-consistently at each filling.
We find that in all cases in a doping range of up to $\simeq20$\% the low-energy physics is dominated by the single active $d_{x^2-y^2}$-orbital, indicated by the roughly linear dependence of the filling of the single active $d_{x^2-y^2}$-orbital in Fig.~\ref{Fig:mainFilling}. 
For the series Nd $\rightarrow$ Sm $\rightarrow$  Y $\rightarrow$ Lu, the range where the 1-band description remains a valid approximation extends further to even larger hole doping.
We additionally note that the self-doping similarly increases by around 4\% along the series from Nd to Lu.  In Fig.~\ref{Fig:mainFilling} this self-doping can be readily identified for the parent compound where for 9 electrons the occupation of the $d_{x^2-y^2}$ orbital is below one (one would correspond to the nominal 3$d^9$ configuration).

The local DMFT spectral functions $A(\omega)$ as well as the self-energies for the 10-orbital model (see SM \cite{SM}) further show that the Ni $3d_{x^2-y^2}$ orbital becomes increasingly less correlated going from the Nd to Lu compound at all doping levels. This is particularly clear for the nominal filling, where Ni $3d_{x^2-y^2}$ in NdNiO$_2$ displays a strongly renormalized quasiparticle peak (with a corresponding self-energy), whereas for LuNiO$_2$ the renormalization is less strong and correspondingly $\Im\Sigma(i\nu_n)$ less peaked (more weakly correlated).

With both the theoretical (DMFT and D$\Gamma$A) and the latest experimental evidence pointing towards a single-band (+electron reservoir) picture, it is justified to downfold to the even simpler effective one-band $d_{x^2-y^2}$ model. The hopping parameters obtained from a projection onto a single, maximally localized Wannier orbital are listed in Table~\ref{Table:hoppings} for the series of compounds considered.
To take into account the self-doping (i.e.~the decoupled electron reservoir) in our downfolding we use the mapping in Fig.~\ref{Fig:mainFilling} between the total doping of the $R$NiO$_2$ nickelate onto that of the  single active Ni 3$d_{x^2-y^2}$ orbital, as done previously in Ref~\cite{Kitatani2020}. 
As discussed above, once approximately 20-25\% doping is reached, the single-orbital model is no longer fully justified for most of the systems considered here, and multiorbital physics sets in.

D$\Gamma$A {\it Results.}
To describe the superconductivity in the system, it is necessary to take into account nonlocal correlations which we do by ladder D$\Gamma$A \cite{Toschi2007,Katanin2009,RMPVertex} using the DGApy implementation of Ref.~\cite{worm2023}. For these  D$\Gamma$A calculations, we consider only the single Wannier orbital model.
For an unbiased prediction, we fix the interaction strength $U$ of this effective 1-band model to those estimated in the original D$\Gamma$A calculation \cite{Kitatani2020} for Sr-doped NdNiO$_2$. This assumes that to a first approximation, only the hopping parameters change when varying the in-plane lattice parameter to that of the various substrates. The local Coulomb interaction is expected to be less sensitive and thus kept constant. 
This leads to an effectively decreasing $U/t$ ratio from Nd to Lu (see discussion below), and is fully consistent with the results of the 10-band model which demonstrated that the $d_{x^2-y^2}$ orbital becomes less correlated along the series, giving further justification that the model results should be representative of the real systems.

We start by calculating the local vertex at DMFT convergence using \textsc{w2dynamics}, then solve the Bethe-Salpeter equation in the particle-hole
channel describing spin and charge fluctuations on an equal footing, and finally take these spin and charge fluctuations  as the irreducible  vertex $\Gamma_{pp}$ (the pairing glue)  in the particle-particle channel
to calculate the superconducting eigenvalues. This is, so to speak, the first step of a more involved full parquet coupling of the various channels. For a review on D$\Gamma$A, see \cite{RMPVertex}; for  specifics on calculating the superconducting $T_c$, see \cite{Kitatani2022}.
The last step, i.e., the particle-particle channel with interacting Green's function $G$ and $\Gamma_{pp}$ corresponds to solving a linearized Eliashberg equation:
\begin{equation}\label{Eq:Delta}
    \lambda\Delta({k})=-\frac{1}{\beta}\sum_{k'}\Gamma_\textrm{pp}(k,k',q=0)G({k'})G(-{k'})\Delta({k'})
\end{equation}
at zero transfer momentum $q=({\bf q},i\omega_m)=({\bf 0},0)$. If  the superconducting eigenvalue $\lambda(T)$ is approaching 1, this indicates the divergence of the superconducting susceptibility.  Thus  $\lambda(T_c)=1$ allows us to identify $T_c$, whereas the gap function $\Delta(k)$ gives the gap symmetry, with a leading $d_{x^2-y^2}$-wave order parameter found for all studied systems. 
For the temperature dependence of $\lambda$ and the fit to extract the critical temperature, see the SM \cite{SM}.

Our calculated $T_c$ for SmNiO$_2$ and LuNiO$_2$  is shown in Fig.~\ref{Fig:Tc_doping} (above) together with the recent experimental data for SmNiO$_2$ from Ref.~\cite{Chow2025}.
The calculated dome is in excellent agreement with experiment both qualitatively and quantitatively
for SmNiO$_2$ with its record $T_c$ among IL nickelates; and we find a maximum calculated  $T_c^\textrm{theory}\approx40$ K that is in remarkable agreement with the experimental $T_c^\textrm{exp}\approx35$ K.
Note that for Sm$_{1-x-y-z}$Eu$_x$Ca$_y$Sr$_z$NiO$_2$ the hole doping is a bit complicated because of the mixed valency of Eu, and we take as ``effective cation doping'' the same
formula as in the experimental paper~\cite{Chow2025,Wei2023}
$\delta=0.6x+y+z$; for the theory, the same total hole doping is modeled as described above.

The inset shows the calculated maximal $T_c$ (at optimal doping) for all systems considered as a function of the in-plane lattice constant of the substrates (the substrates are indicated on the upper abscissa).
For the full phase diagrams of the other IL nickelates considered, see the SM \cite{SM}.
Experimental results using the corresponding substrates and the previous D$\Gamma$A calculations for NdNiO$_2$ \cite{Kitatani2020} (which used DFT relaxed bulk lattice parameters) are also shown.

\begin{figure}[t]
	\centering
    \includegraphics[angle=0, width=1.0 \columnwidth]{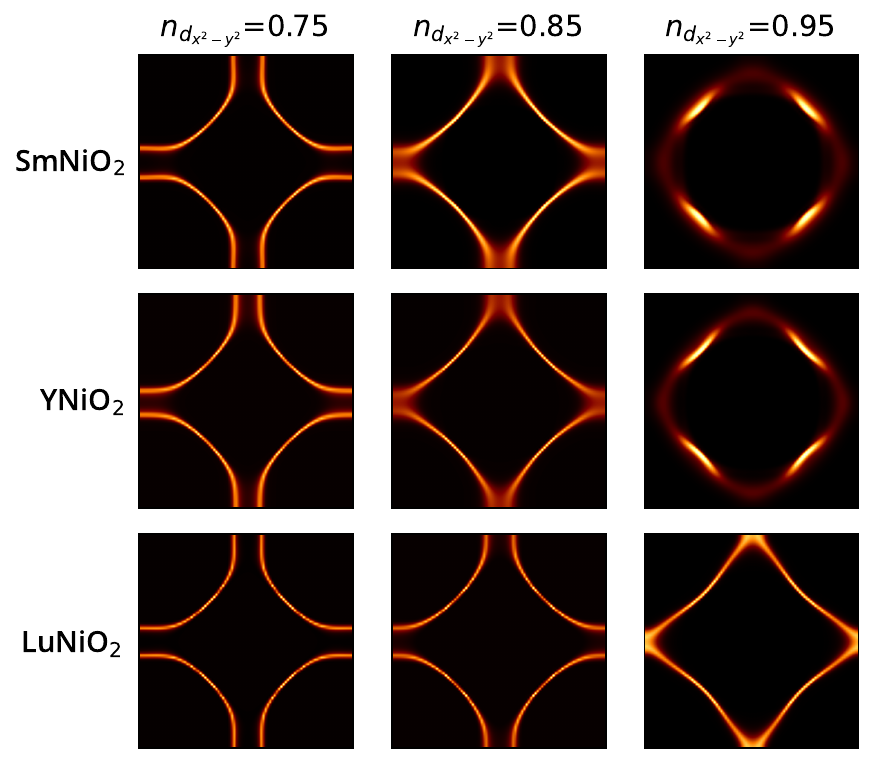}
    \caption{\label{Fig:FS}
    Fermi surfaces calculated with D$\Gamma$A at 58 K ($\beta=200$ eV$^{-1}$) for $n_{x^2-y^2}=0.75$, $0.85$ and $0.95$ in the $k_z=0$ plane.
    Top row: SmNiO$_2$@NGO, Middle row: YNiO$_2$@LAO, and Bottom row: LuNiO$_2$@YAP. Note that $n_{d_{x^2-y^2}}=0.75$ is not reached when doping SmNiO$_2$ since holes instead start to empty  other 3$d$ orbitals in this doping range, see Fig.~\ref{Fig:mainFilling}.  The smaller $U/t$ ratio in LuNiO$_2$ does not lead to the emergence of a pseudogap any more, unlike in the Nd- (not shown), Sm- and Y-based compounds.}
\end{figure}

Fixing the in-plane lattice parameter to that of the substrate (i.e., assuming epitaxial films) leads to an effective decrease of  $U/t$ from
Nd$\rightarrow$Sm$\rightarrow$Y$\rightarrow$Lu. At first, from Nd to Y, $T_c$ increases slightly; and the  region where superconductivity is found is shifted to lower dopings, see SM~\cite{SM}. Eventually the phase diagram
changes  qualitatively and more dramatically with $T_c$ no longer suppressed  at low doping for LuNiO$_2$. To understand this behavior, let us recall that
the superconducting dome in Nd(Sm)NiO$_2$ emerges from a competition of two effects:
(i) Starting from the overdoped regime and reducing hole doping, antiferromagnetic spin fluctuations are enhanced, thus $\Gamma_{pp}$ in Eq.~(\ref{Eq:Delta}) becomes larger, and consequentially $T_c$ increases (from right to left in Fig.~\ref{Fig:Tc_doping}) . (ii) However, at some point
spin fluctuations are so large that a pseudogap opens. This means that the Green function lines $G$ in Eq.~(\ref{Eq:Delta}) are damped (become decoherent), and thus
eventually $T_c$ decreases again in the underdoped regime \cite{NoteFK}. This pseudogap for SmNiO$_2$ is shown in Fig.~\ref{Fig:FS} (top right).

Moving on to the hypothetical LuNiO$_2$  film with the $a$ lattice parameter fixed by YAP,
this pseudogap is suppressed, see Fig.~\ref{Fig:FS} (bottom). As a consequence, $T_c$ continues to increase in the low doping regime. There is no longer a superconducting dome, and much larger $T_c$'s are thus possible.

Let us note, however, that at very small doping levels a competing antiferromagnetic order is possible, which can have a  pairbreaking effect, thus possibly recovering a dome-like $T_c$ curve but with the downturn now for much smaller dopings. The regime where a prospective antiferromagnetic order could possibly occur is indicated in Fig.~\ref{Fig:Tc_doping} as the gray-shaded area. Let us note that so far in experiment on IL nickelates no antiferromagnetic order was found. Partially this can be explained by the self-doping due to the $R$ electron pockets. This self-doping leads to $\sim 5$-$6\%$ hole doping even in the absence of chemical doping, cf.~Fig.~\ref{Fig:mainFilling}. It is worth to note that, as we show here, this self-doping varies somewhat between the different IL nickelates. For such a (self-)doping level, also many cuprates do not show long-range antiferromagnetic order any longer.

{\it Summary.}
Our calculations for the superconducting phase diagram of the recently synthesized SmNiO$_2$ with a record $T_c$ show excellent agreement with experiment \cite{Chow2025}.
The observed larger $T_c$ can be explained on the one hand by the smaller $U/t$ ratio due to the --compared to earlier experiments of NdNiO$_2$ on STO-- compressed in-plane lattice constant.
However, on the other hand, the quality of the (difficult to synthesize) films remains an important aspect when comparing different IL nickelates.
The much lower resistivity of the newly synthesized SmNiO$_2$ films \cite{Chow2025} (which cannot be explained by the smaller lattice constant)
indicates much less stacking faults and disorder than in other
nickelate films. This effect is beyond the realms of our calculations, but is certainly important in order to get larger $T_c$'s in experiment and subsequently to come closer to our theoretical estimates.
Beyond the hitherto experimentally synthesized films, we consider YNiO$_2$ and  LuNiO$_2$ on LAO and YAP substrates. Our prediction is that there is still plenty of room to further
enhance $T_c$  quite dramatically when further reducing the in-plane lattice constants through suitably chosen substrates.

{\it Acknowledgments.}\\
We thank Ariando, Eric Jacob, Oleg Janson, Leonard Verhoff,  Wenfeng Wu and especially Liang Si for inspiring discussions.
We further acknowledge funding by the Austrian Science Funds (FWF) through project DOI 10.55776/I5398. The DFT+DMFT calculations have been mainly done on the Vienna Scientific Cluster (VSC).

For the purpose of open access, the authors have applied a CC BY public copyright license to any Author Accepted Manuscript version arising from this submission.


\bibliography{main,main2}%

\begin{thebibliography}{44}
\bibitem [{pse()}]{pseudocubic}%
  \BibitemOpen
  \href@noop {} {}\bibinfo {note} {The substrates (except for STO) and parent compounds $R$NiO$_3$ adopt distorted perovskite structures, and we use the corresponding pseudocubic lattice parameters for the strain estimates and the epitaxially fixed in-plane ($a=b$) lattice constants.}\BibitemShut {Stop}%
\end{thebibliography}

\end{document}


\title{Supplemental Material: Substrate and cation engineering for \\optimizing superconductivity in infinite-layer nickelates}
\author{Viktor Christiansson\orcidlink{0000-0002-9002-0827}}
\affiliation{Institute of Solid State Physics, TU Wien, 1040 Vienna, Austria}

\author{Karsten Held\orcidlink{0000-0001-5984-8549}}
\affiliation{Institute of Solid State Physics, TU Wien, 1040 Vienna, Austria}

\begin{abstract}
Here we provide supplemental information that support our calculations and make our results better reproducible:
In Section~\ref{Sec:computational} we detail the computational methods used in the present work.
Section~\ref{Sec:DFT} shows the density functional theory (DFT) bandstructure and its 10- and one-band Wannierization.
Section~\ref{Sec:lambda} shows the inter- and extrapolation of the superconducting eigenvalue $\lambda$.
In Section~\ref{Sec:Tc}, we provide the phase diagram, critical temperature vs.~doping, for all compounds studied.
Finally, Section~\ref{Sec:DMFT} presents the local dynamical mean-field theory (DMFT) spectral functions akin to the density of states and the DMFT self-energy.
\end{abstract}
\maketitle

\section{Computational details}
\label{Sec:computational}
As a starting point, we perform density functional theory (DFT) calculations for infinite-layer (IL) nickelates $R$NiO$_2$ ($R=$Nd, Sm, Y, Lu) in the $P4/mmm$ space group at fixed in-plane lattice parameters $a=b$, dictated by the substrate considered (see the main text). We use a $24\times24\times24$ {\bf k}-grid, and we employ the gradient approximation (GGA) with Perdew-Burke-Ernzerhof (PBE) exchange correlation functional \cite{Perdew96} as implemented in the \textsc{WIEN2k} code \cite{blaha2001wien2k}. Where applicable, we consider the $R$ $4f$-states as frozen in the core and do not take these into account in the valence window, in line with the standard treatment of these systems.
Since we assume that the substrate fixes the in-plane lattice parameters, the out-of-plane $c$ lattice constant is the only remaining parameter governing the structure. We compute this by performing a DFT calculation for a range of $c$ values and using a fit to a Birch-Murnaghan equation of state.

We want to consider the low-energy physics and define a model subspace that captures the processes of interest. Using the relaxed $c$ parameters for a given substrate (i.e. $a=b$ in-plane lattice parameters), we define a 10-band model by projecting onto maximally localized Wannier functions on the Ni and $R$ $d$ states using the
\textsc{wien2wannier} \cite{Kunes2010a} and \textsc{Wannier90} \cite{Pizzi2020} codes. 
The DFT band structures and Wannier tight-binding models are shown in Sec.~\ref{Sec:DFT}.

Since we are considering a model with correlated (localized) $d$-states, we next perform a DFT+dynamical mean-field theory (DMFT) calculation for this 10-band model at inverse temperature $\beta=40$ eV$^{-1}$ ($T=290$ K) using the hybridization-expansion continuous-time quantum Monte Carlo (CT-QMC) code \textsc{w2dynamics} \cite{w2dynamics2018}. 
We consider two impurity problems, with the five orbitals centered on the Ni and $R$ sites, respectively. 
The impurities we consider have atom-dependent local static Kanamori interactions: $U_\textrm{Ni}=4.4$ eV, $J_\textrm{Ni}=0.65$ eV, $V_\textrm{Ni}=U-2J=3.1$ eV for the Ni site, and $U_{R}=2.5$ eV, $J_{R}=0.25$ eV, $V_{R}=U-2J=2.0$ eV for the $R$ site, to be consistent with typical parameters used in the literature. While these values are of course in-principle dependent on both the strain and $R$ for the considered IL nickelate \cite{Petocchi2020,Si2020,Christiansson2023}, we instead keep them fixed to that of previous calculations~\cite{Si2020}. This choice clearly avoids to have any possibility of ``fine-tuning" the interaction.

An alternative method would be to do a full constrained random-phase approximation for each system. As we are only considering a static interaction, which would typically be considered as the static part of the so-calculated interaction $U^\textrm{cRPA}(0)$. This would as a matter of course come with it's separate problem of underestimating the local effective interaction (since we are then missing the dynamical screening, which at higher energy is less efficient and the interaction asymptotically goes to the bare Coulomb interaction). What is typically done is therefore to adjust up the interaction strength manually to take this into account.
The strain and $R$-dependence  of the interaction is, in any case, 
not significant enough to change our conclusions that the increased in-plane strain is mainly responsible for the significant increase in $T_c$ observed experimentally).

In the self-consistent solution of the DMFT impurity problem, we use $\sim3\times10^8$ measurements in each iteration and the calculations require on the order of $30$-$40$ iterations for convergence.
To mimic the effect of the chemical doping experimentally realized by doping with Sr/Ca/Eu, we perform calculations at varying filling of the model (the nominal filling in the undoped compound is 9 electrons), where the chemical potential is updated self-consistently to match.

\begin{figure}[b]
	\centering
\includegraphics[angle=0, width=0.41 \columnwidth]{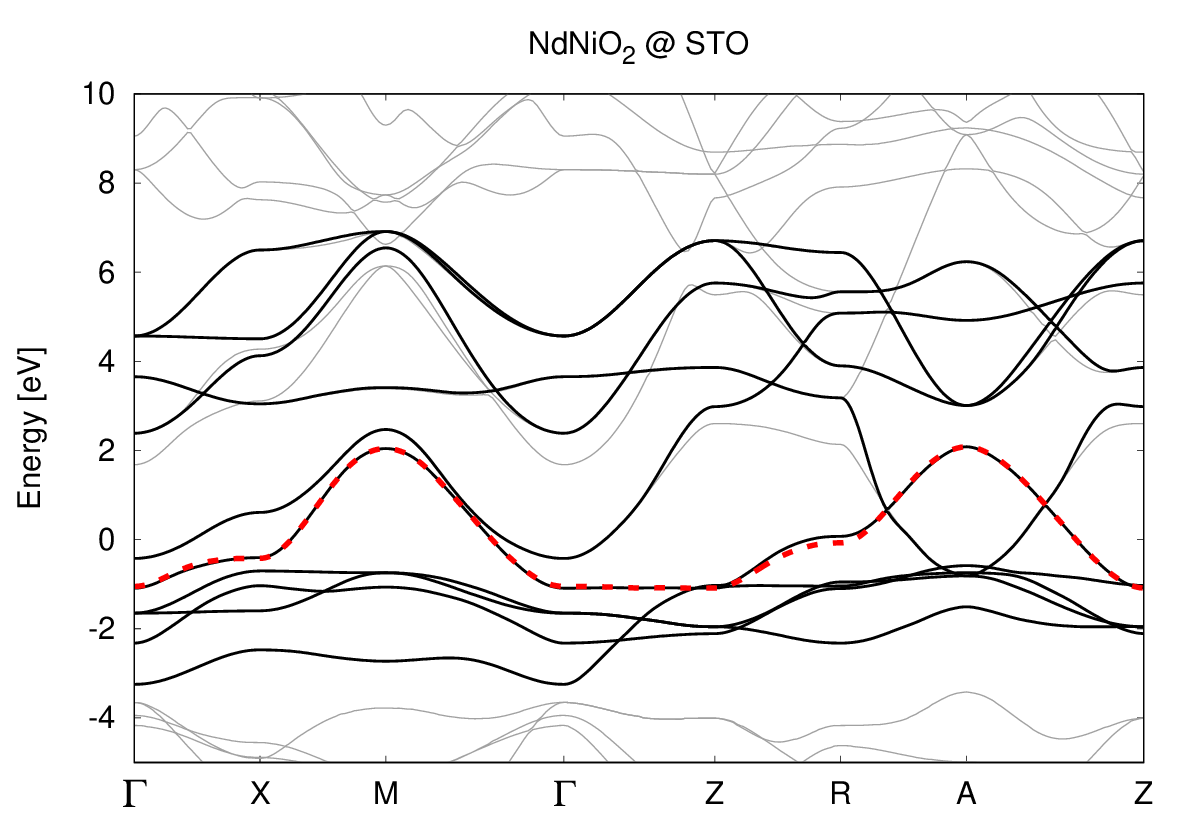}
\includegraphics[angle=0, width=0.41 \columnwidth]{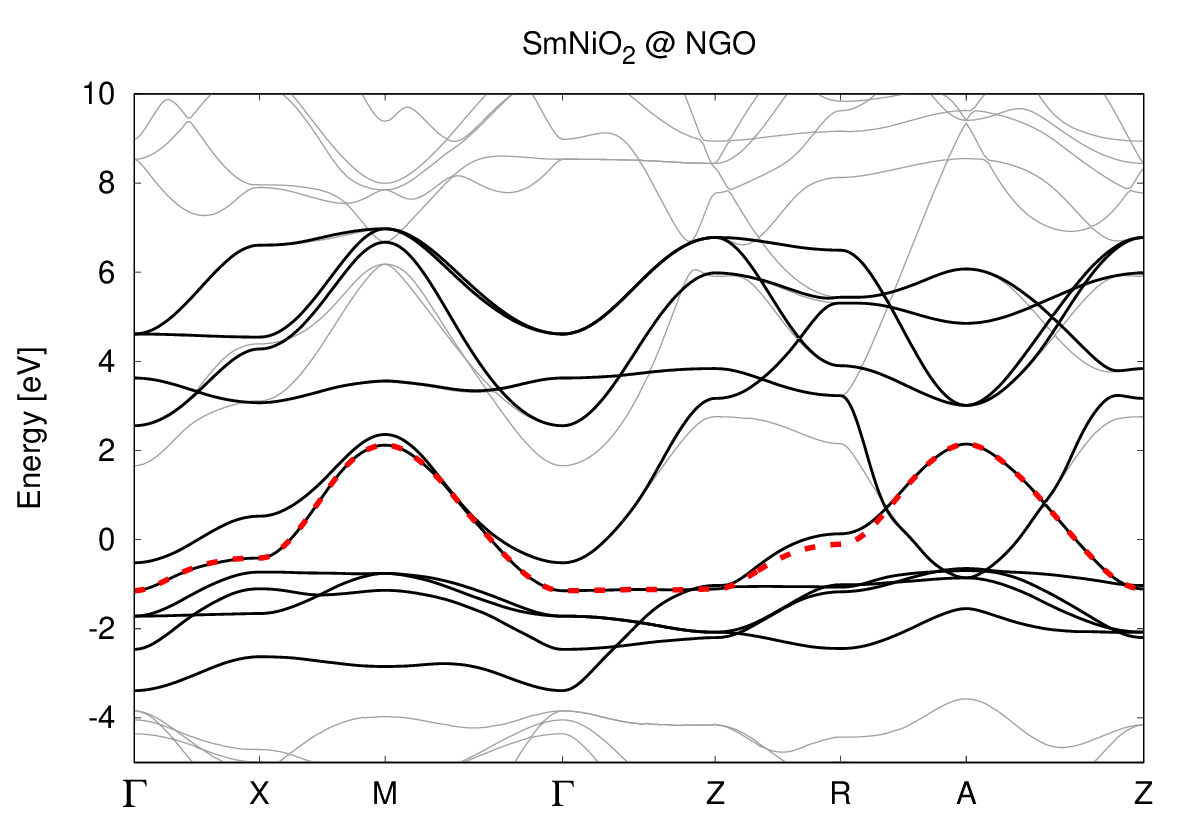}
\includegraphics[angle=0, width=0.41 \columnwidth]{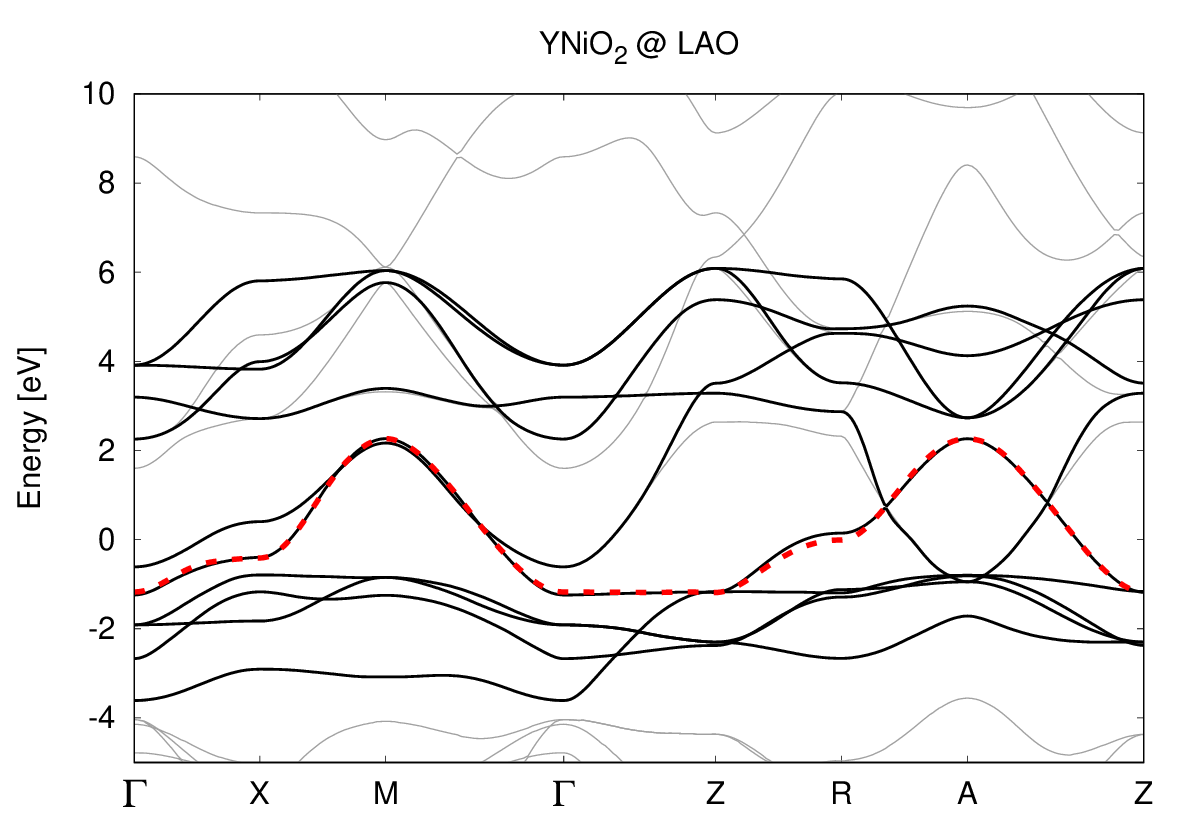}
\includegraphics[angle=0, width=0.41 \columnwidth]{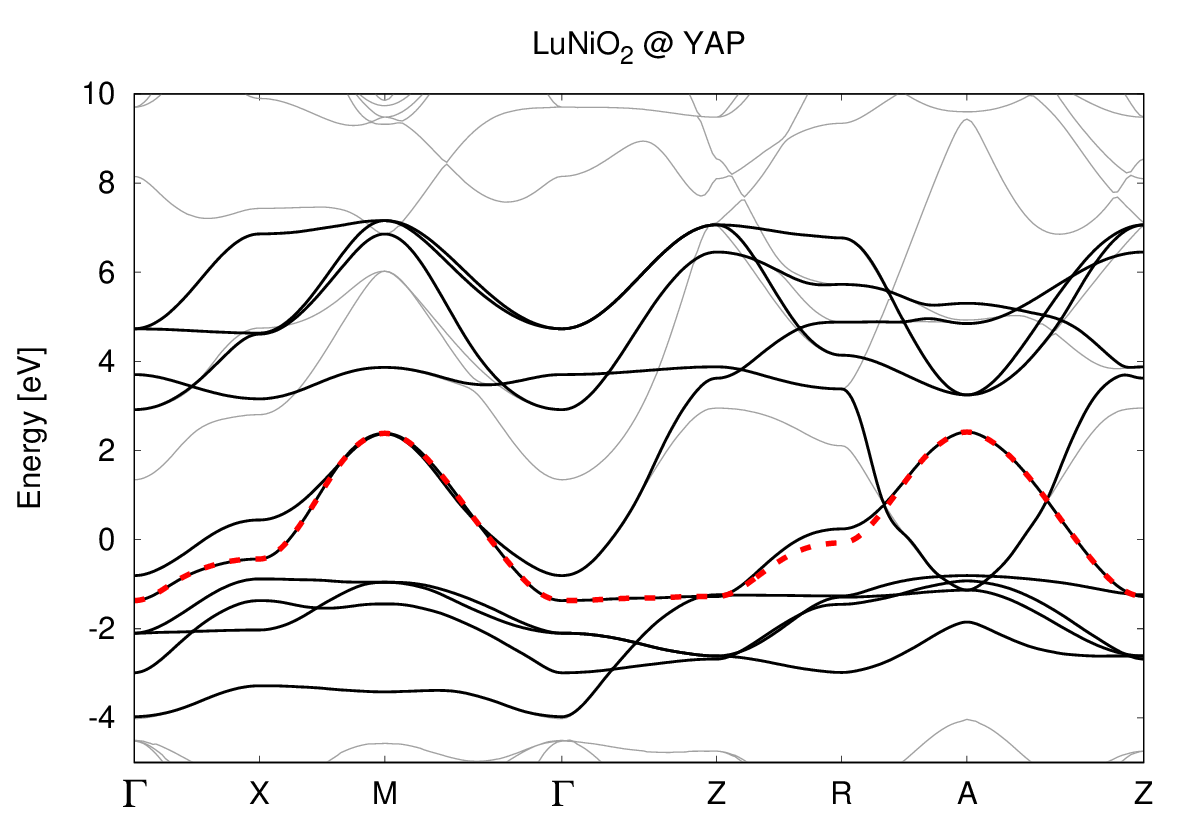}
\caption{\label{Fig:DFT_BS} 
DFT calculated band structure (gray lines) along the indicated high-symmetry path. The 10-band model is shown by the thicker black lines, and the 1-band model by the dashed red line. Top left: NdNiO$_2$ on STO, top right:  SmNiO$_2$ on NGO, bottom left:  YNiO$_2$ on LAO, bottom right:  LuNiO$_2$ on YAP.
}
\end{figure}

As the low-energy physics is dominate by the Ni $3d_{x^2-y^2}$ state in a large doping range, we also define a second model with this single orbital by similarly projecting onto maximally localized Wannier functions (see Fig.~\ref{Fig:DFT_BS} for the Wannier projections of the different systems considered).
From this model, we extract up to the third nearest neighbour hopping parameters ($t$, $t'$ and $t''$).
We note here that these hopping parameters are very insensitive to the out-of-plane $c$ lattice parameter, and varies only within $<$0.5\% (in most cases much less) when the $c$-lattice parameter is increased by $>$5\%. Since the effective interaction-to-hopping ratio $U/t$ is found to be the main ingredient in the observed increase in $T_c$, our results are not sensitive to the DFT relaxation, as long as we assume that we consider the substrate to (epitaxially) fix the in-plane lattice parameters.
It is these in-plane lattice parameters to which the $t$, $t'$ and $t''$ hopping parameters are sensitive.

We next perform DMFT calculations at different fillings (taking into account both the intrinsic self-doping and the extrinsic doping, see the main text) and temperatures down to 46 K  ($\beta=250$ eV$^{-1}$) for an interaction $U=3.2$ eV (with a similar consideration for the fixed value of $U$ as above). For the CT-QMC sampling we use an increasing number of measurements at low temperature, and we find a well-converged solution at the lowest temperature (46 K) using $10^9$ measurements for the one-particle quantities and a $200\times200$ {\bf k}-grid for an accurate Brillouin zone integration after $\sim30$ iterations.

From this converged solution, we perform a higher statistics run with $\sim 4 \times 10^9$ measurements with worm sampling \cite{Gunacker2015} to obtain 1-particle quantities with minimal noise. We then calculate the 2-particle Green's function $G^{(2)}$ for this impurity problem and from this extract the local 2-particle vertex in the particle-hole channel $\Gamma^{ph}$. This together with the 1-particle quantities obtained in the previous step forms the basic ingredients needed for calculating the nonlocal suseptibility $\chi(\mathbf{q},i\omega_m)$ and self-energy $\Sigma(\mathbf{k},i\nu_n)$ within the ($\lambda$-corrected) ladder dynamical vertex approximation (D$\Gamma$A).
For details on the computational steps involved, see e.g. Refs.~\cite{Worm2024,Rohringer2018b,Kitatani2022}

For the D$\Gamma$A calculations we use the DGApy code \cite{worm2023} and a momentum grid of $96\times96$ ($120\times120$ at higher temperatures). The code treats the frequency grid in different ``boxes" with a core region where the calculated impurity vertex is used (here we have used up to 180 Matsubara frequencies at the lowest temperature) and a shell region for vertex asymptotics (up to 300 Matsubara frequencies), for better convergence.
For additional details on the treatment, we refer the reader to Ref.~\cite{Kitatani2022}.
Finally, we solve the linearized Eliashberg equation in the particle-particle channel using the interacting 1-particle Green's function $G$ and $\Gamma_{pp}$ (nonlocal vertex in the particle-particle channel):
\begin{equation}\label{Eq:Delta}
    \lambda\Delta({k})=-\frac{1}{\beta}\sum_{k'}\Gamma_\textrm{pp}(k,k',q=0)G({k'})G(-{k'})\Delta({k'})
\end{equation}
at zero transfer momentum $q=({\bf q},i\omega_m)=({\bf 0},0)$. An eigenvalue $\lambda$ that crosses 1 indicates an instability in the particle-particle channel, i.e. a superconducting instability. This criterion is used to calculate the critical temperature. For the explicit calculated $\lambda(T)$ and the fits used to obtain the $T_c$'s reported in the main text, see Sec.~\ref{Sec:lambda}.

\section{DFT band structure and Wannier model bands}
\label{Sec:DFT}

In Fig.~\ref{Fig:DFT_BS} we show the DFT band structures as well as the 10-band Wannier model and the single-band effective model used in the D$\Gamma$A $T_c$ calculations for NdNiO$_2$ @ STO, SmNiO$_2$ @ NGO, YNiO$_2$ @ LAO, and LuNiO$_2$ @ YAO.

\section{Eigenvalues of the linearized Eliashberg equation}
\label{Sec:lambda}
\begin{figure}[h]
	\centering
\includegraphics[angle=0, width=0.4 \columnwidth]{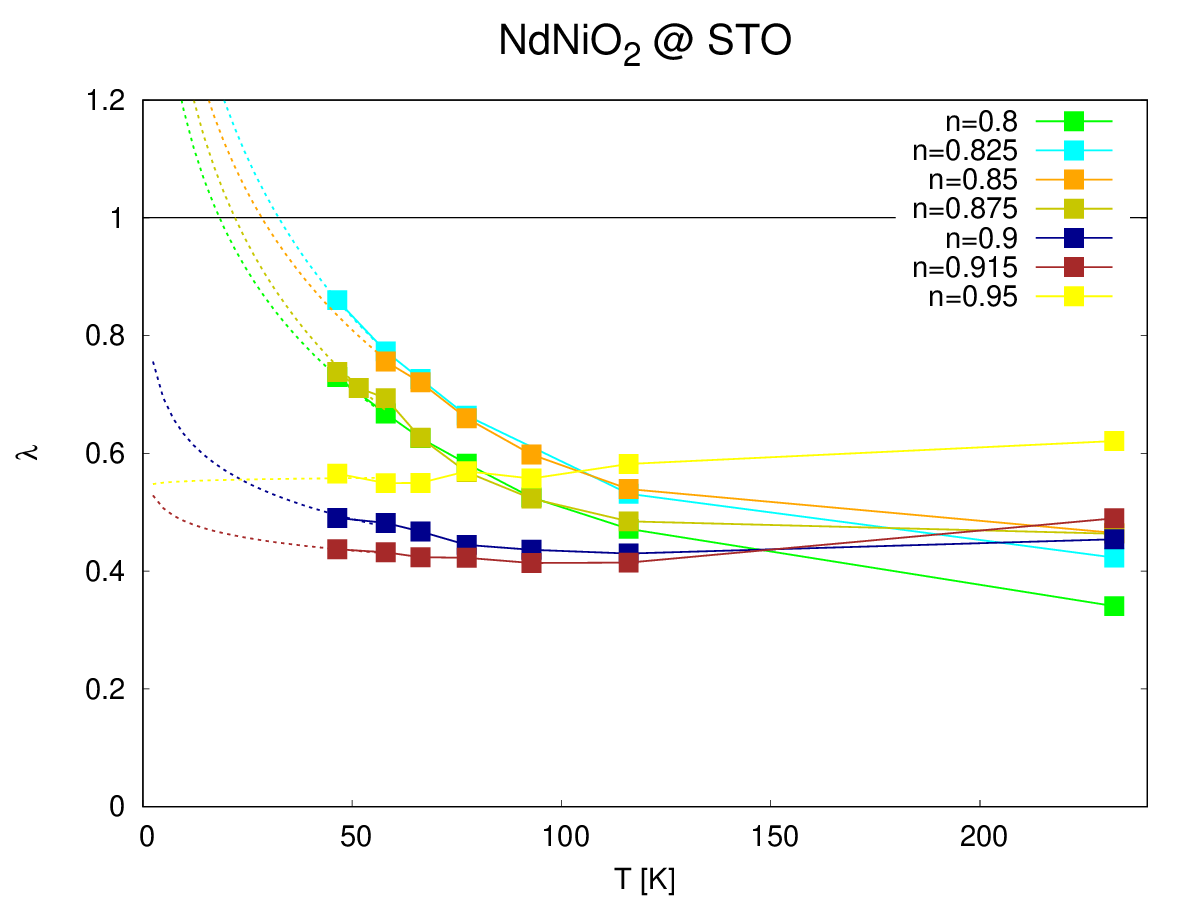}
\includegraphics[angle=0, width=0.4 \columnwidth]{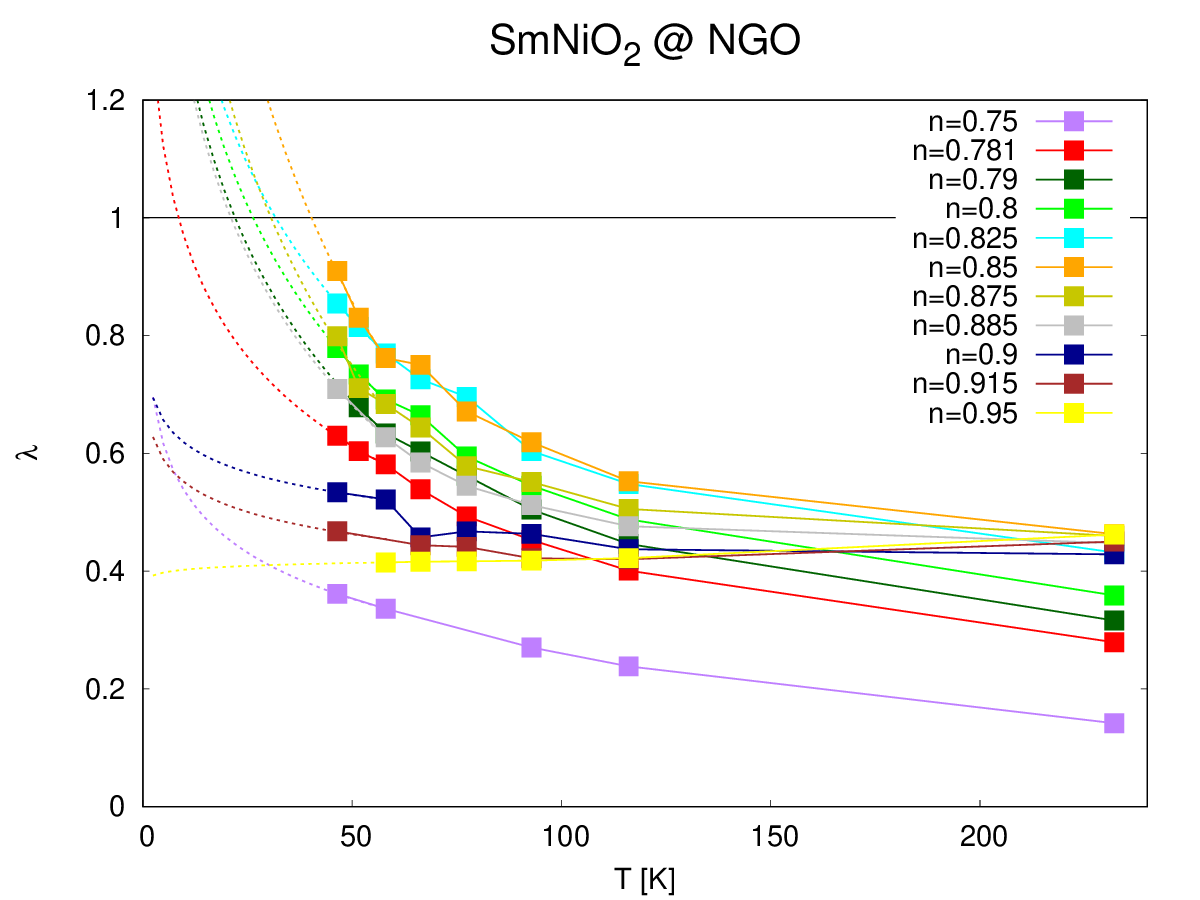}
\includegraphics[angle=0, width=0.4 \columnwidth]{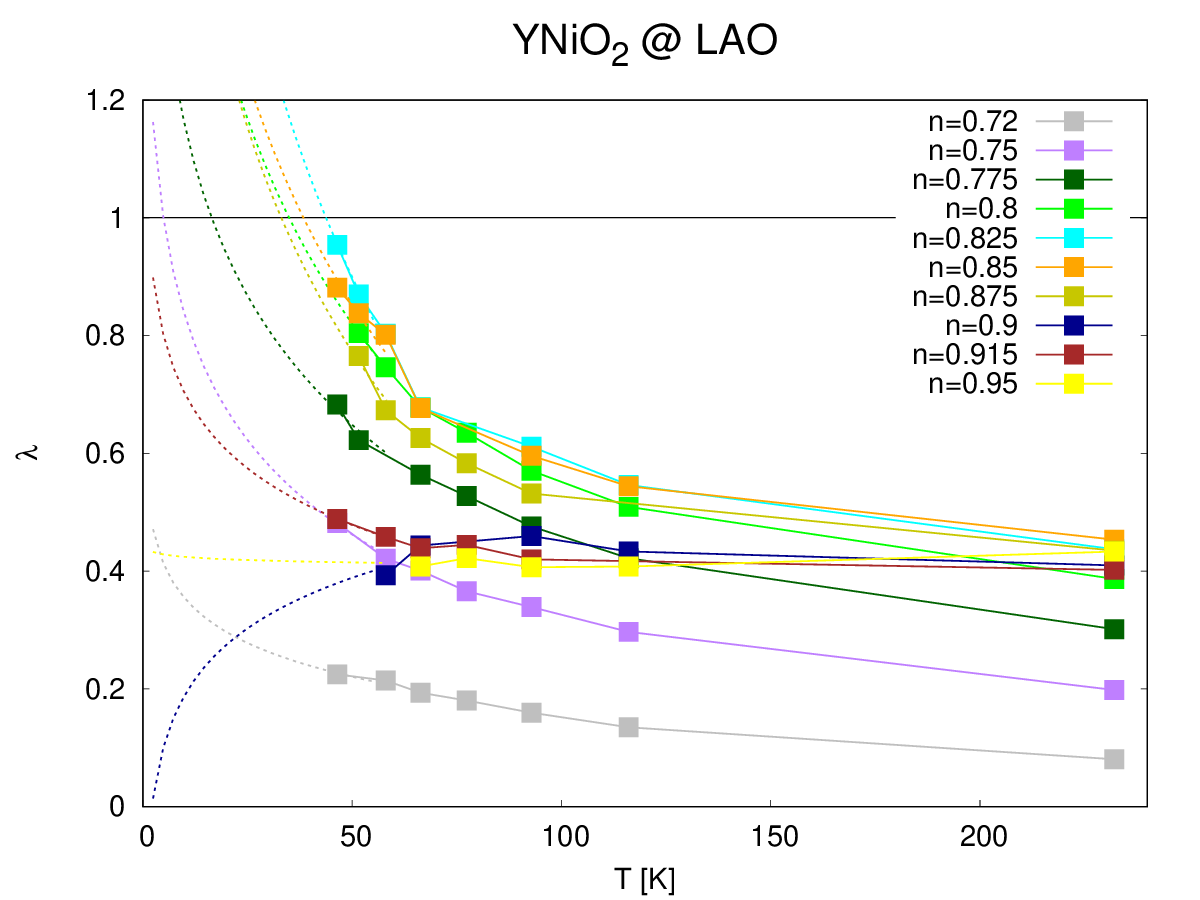}
\includegraphics[angle=0, width=0.4 \columnwidth]{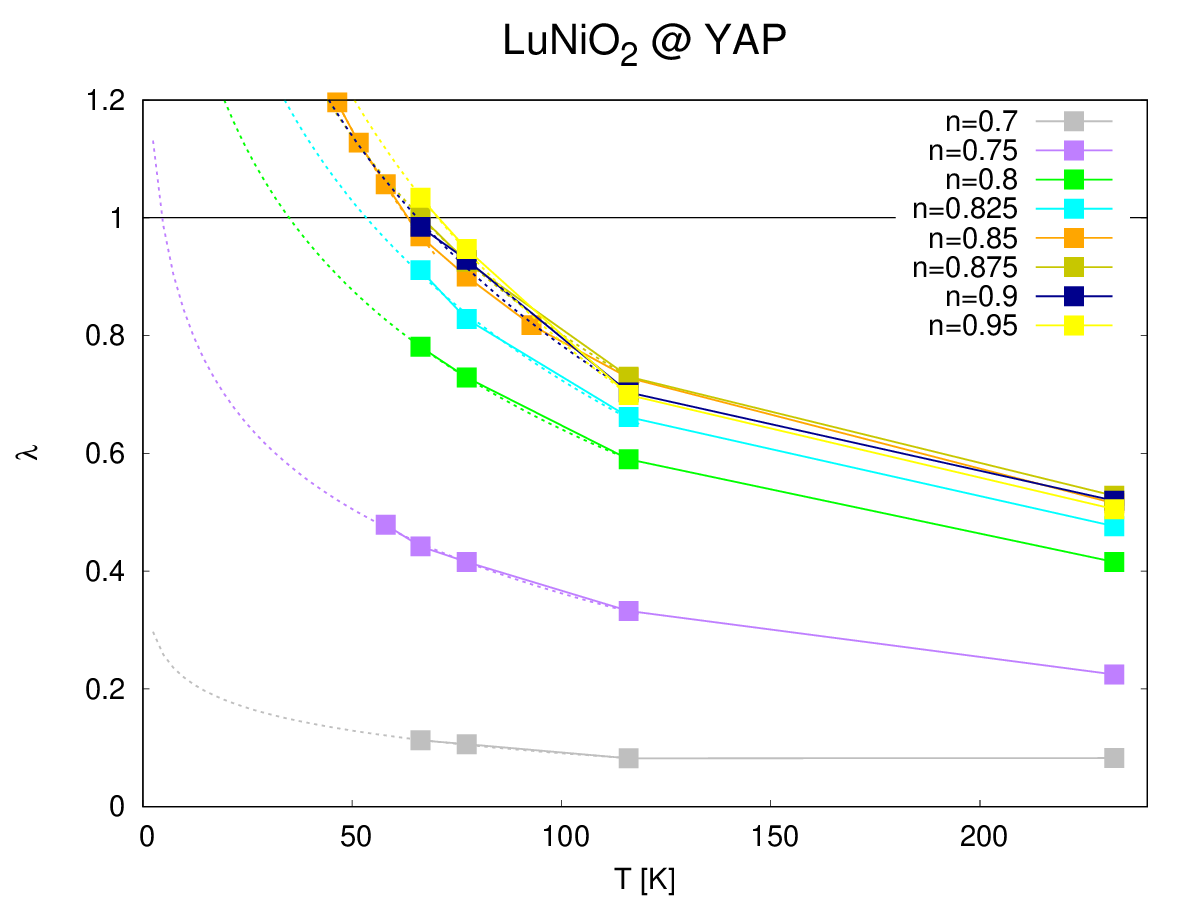}
\caption{\label{Fig:Eliashberg_eigenvalues}
Eigenvalues $\lambda(T)$ for NdNiO$_2$ on STO (top left) SmNiO$_2$ on NGO (top right), YNiO$_2$ on LAO (bottom left), and LuNiO$_2$ on YAP (bottom right) obtained from solving the linearized Eliashberg equation, Eq.~\eqref{Eq:Delta}, together with a low-temperature fit of the form $\lambda(T)=A-B\ln\left( 1/T \right)$ (dashed lines). $\lambda(T)=1$ indicates an instability towards a superconducting order and is used to determine the critical temperature.
}
\end{figure}

As described in the computational details section above, and in the main text, the superconducting critical temperature is calculated by solving the linearized Eliashberg equation, Eq.~\eqref{Eq:Delta}, and determining where the eigenvalue $\lambda$ crosses 1. In the cases where we are unable to reach the temperatures needed to observe this directly, we fit the low-temperature data points using a function $\lambda(T)=A-B\ln\left( 1/T \right)$. The calculated eigenvalues $\lambda(T)$ are reported in Fig.~\ref{Fig:Eliashberg_eigenvalues} (solid squares) together with the low-temperature fits (dashed lines).

\section{$T_c$ Phase diagram for the considered $R$NiO$_2$ compounds}
\label{Sec:Tc}

\begin{figure}[h]
	\centering
\includegraphics[angle=0, width=0.45 \columnwidth]{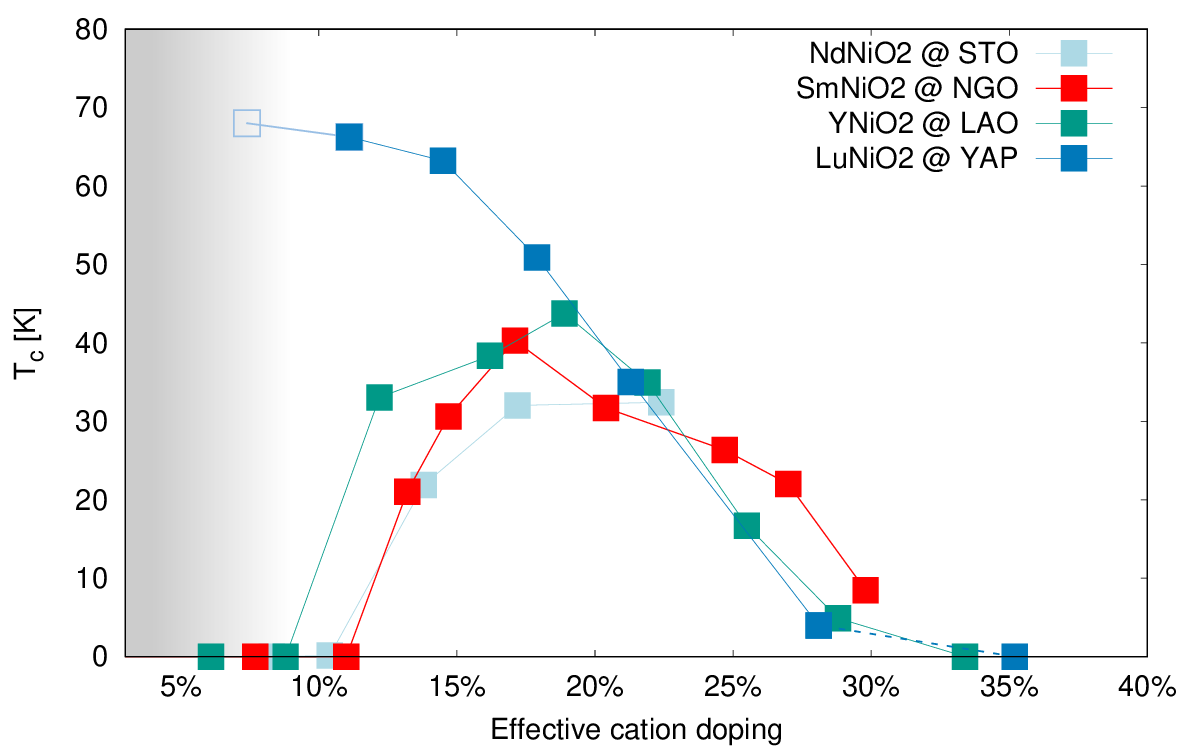}
\caption{\label{Fig:PDall}
$T_c$ vs.\ effective doping of experimentally realized NdNiO$_2$, SmNiO$_2$, hypothetical YNiO$_2$ and LuNiO$_2$.
The shaded area indicates a region where a competing antiferromagnetic order could be expected, see the main text. 
}
\end{figure}

The full $T_c$ phase diagrams of the the calculated IL nickelates (NdNiO$_2$, SmNiO$_2$, YNiO$_2$, and LuNiO$_2$) is shown in Fig.~\ref{Fig:PDall}. For NdNiO$_2$ and SmNiO$_2$ we do not show further dopings in the overdoped regime, as at these doping levels multi-orbital physics becomes relevant. Further doping within the one-band model for these systems (not shown),  leads to the expected suppression of $T_c$ giving the full dome behavior with no superconductivity on the overdoped side as seen experimentally.

\pagebreak
\section{10-band DMFT spectral function and self-energy}
\label{Sec:DMFT}
\begin{figure}[t]
	\centering
    \vspace{-1.2cm}
\includegraphics[angle=0, width=0.54 \columnwidth]{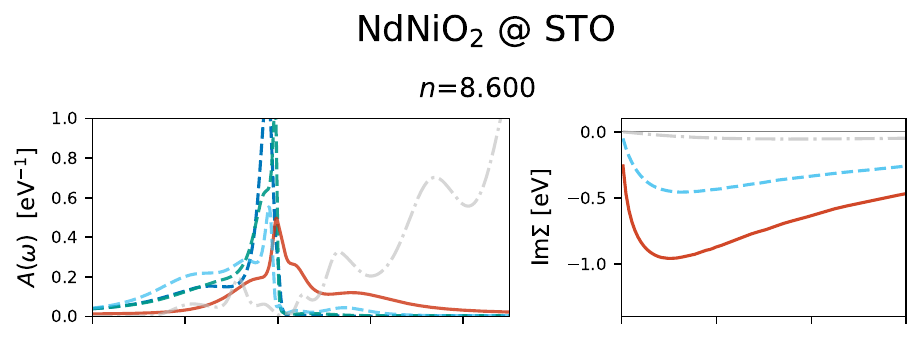}
\includegraphics[angle=0, width=0.54 \columnwidth]{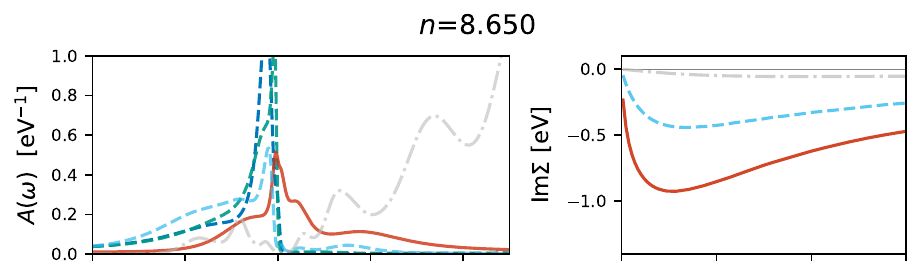}
\includegraphics[angle=0, width=0.54 \columnwidth]{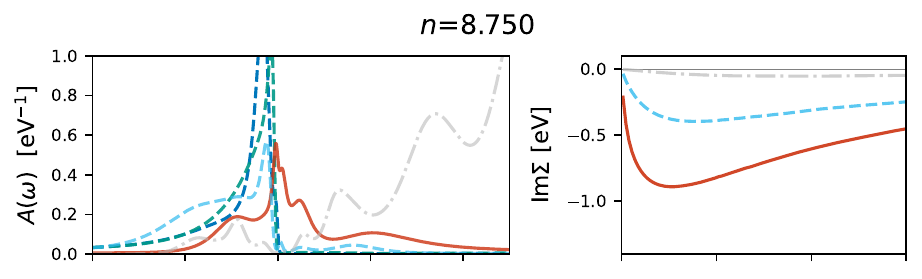}
\includegraphics[angle=0, width=0.54 \columnwidth]{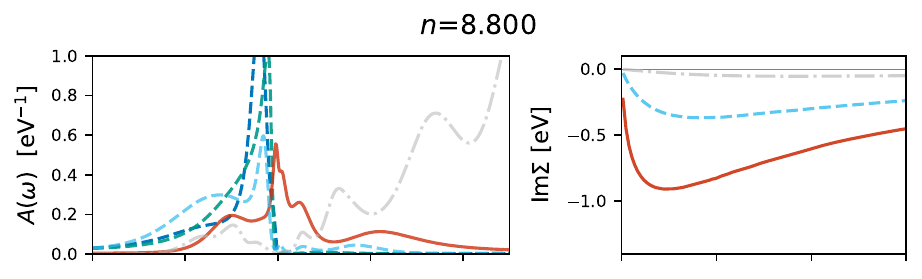}
\includegraphics[angle=0, width=0.54 \columnwidth]{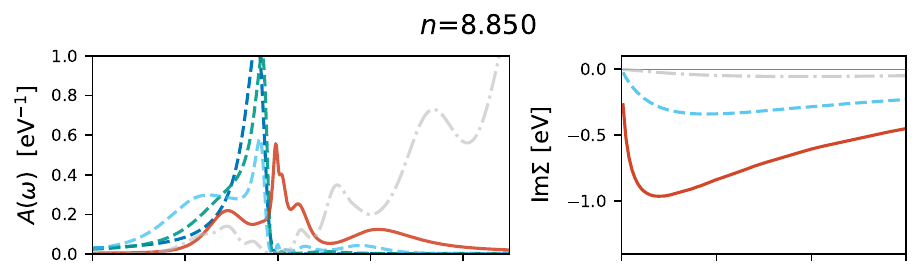}
\includegraphics[angle=0, width=0.54 \columnwidth]{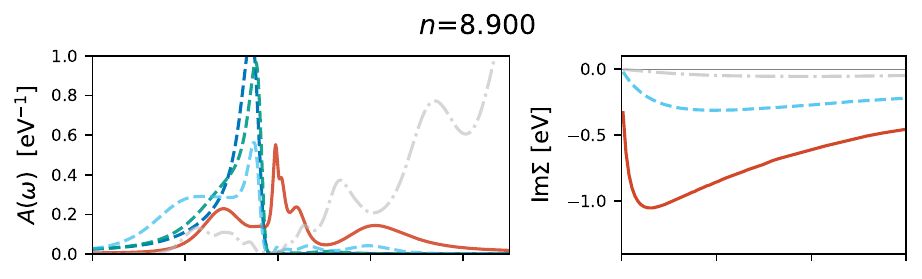}
\includegraphics[angle=0, width=0.54 \columnwidth]{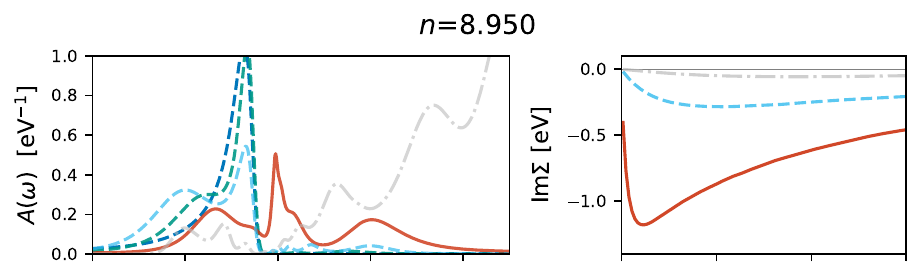}
\includegraphics[angle=0, width=0.55 \columnwidth]{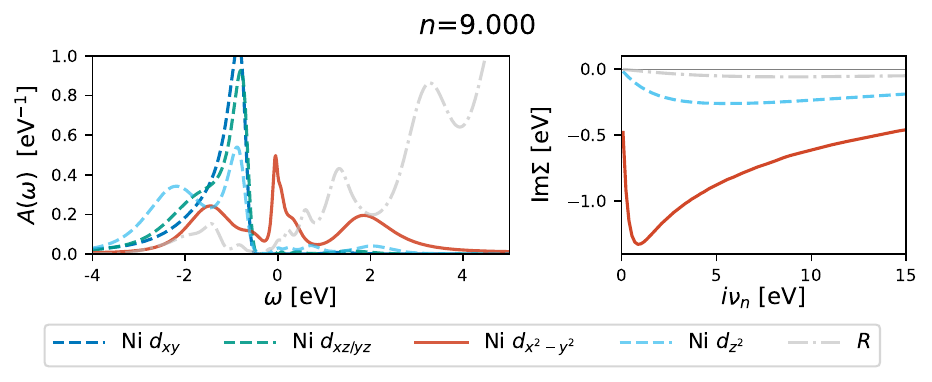}

\caption{\label{Fig:Nd_spectra}
Left:  Local DMFT spectral function for NdNiO$_2$@STO at different doping levels (the nominal filling is 9 electrons) for the 10-orbital model. Right: the DMFT self-energy of the representative Ni $3d_{z^2}$, Ni $3d_{x^2-y^2}$, and Nd $5d_{z^2}$ orbitals.
}
\end{figure}

\begin{figure}[t]
	\centering
    \vspace{-1.2cm}
\includegraphics[angle=0, width=0.54 \columnwidth]{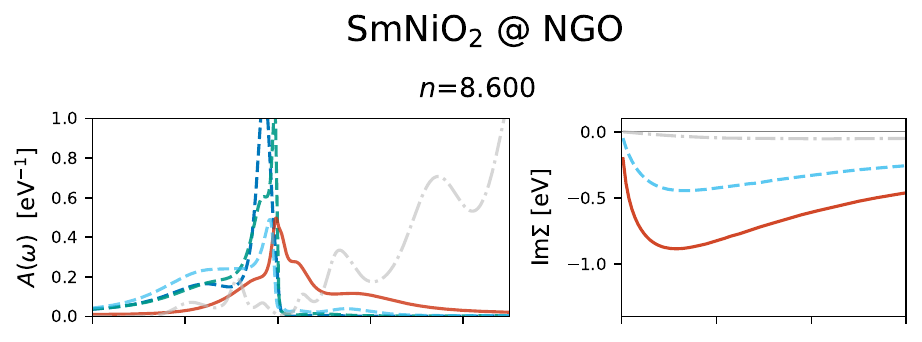}
\includegraphics[angle=0, width=0.54 \columnwidth]{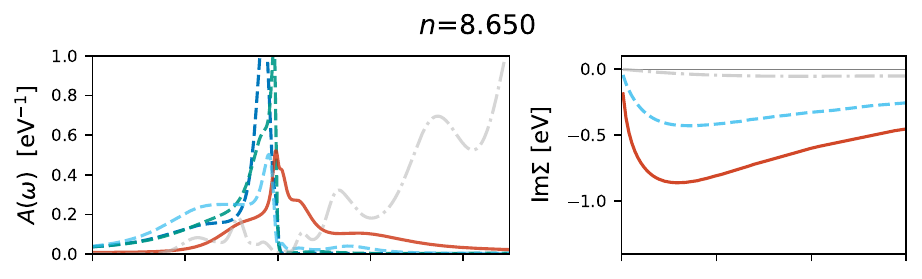}
\includegraphics[angle=0, width=0.54 \columnwidth]{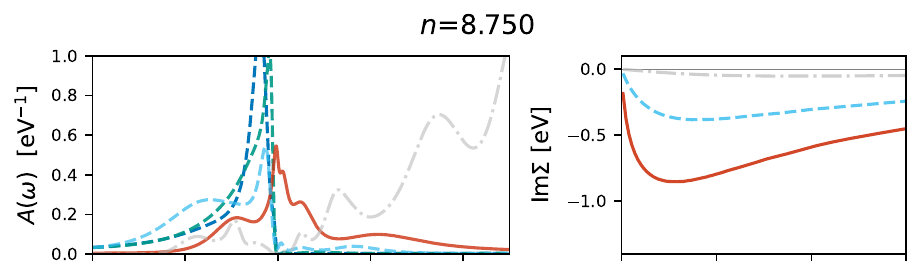}
\includegraphics[angle=0, width=0.54 \columnwidth]{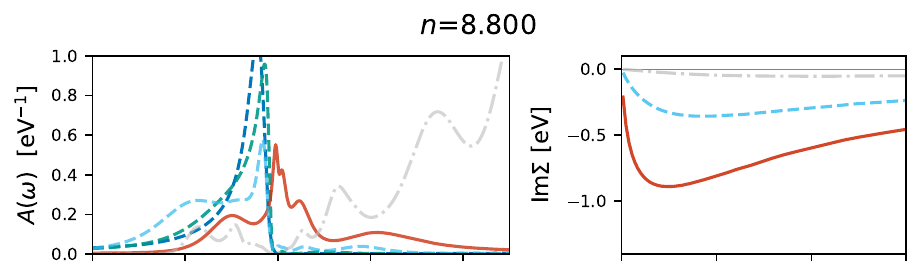}
\includegraphics[angle=0, width=0.54 \columnwidth]{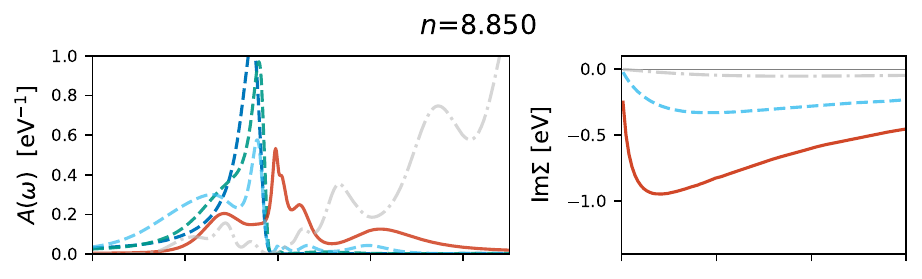}
\includegraphics[angle=0, width=0.54 \columnwidth]{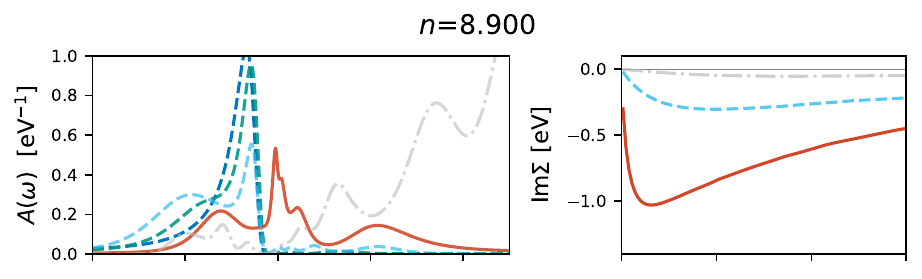}
\includegraphics[angle=0, width=0.54 \columnwidth]{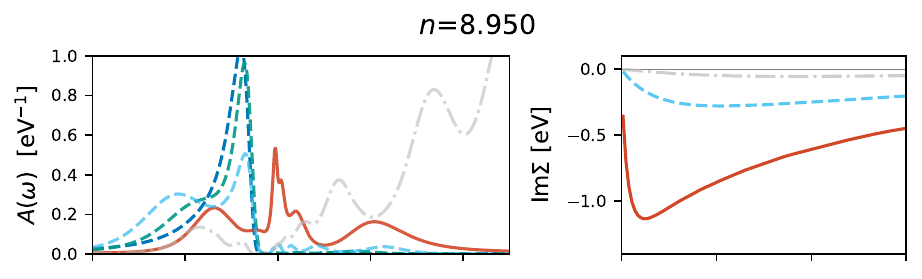}
\includegraphics[angle=0, width=0.55 \columnwidth]{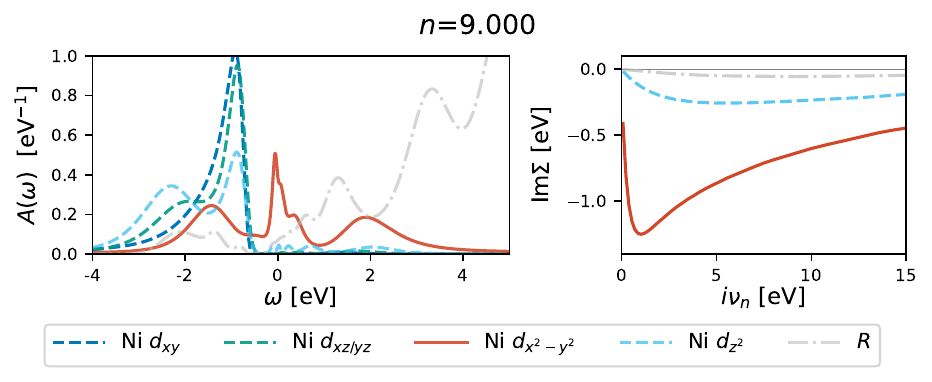}

\caption{\label{Fig:Sm_spectra}
Same as Fig.~\ref{Fig:Nd_spectra} for SmNiO$_2$@NGO.
}
\end{figure}

\begin{figure}[t]
	\centering
    \vspace{-1cm}
\includegraphics[angle=0, width=0.54 \columnwidth]{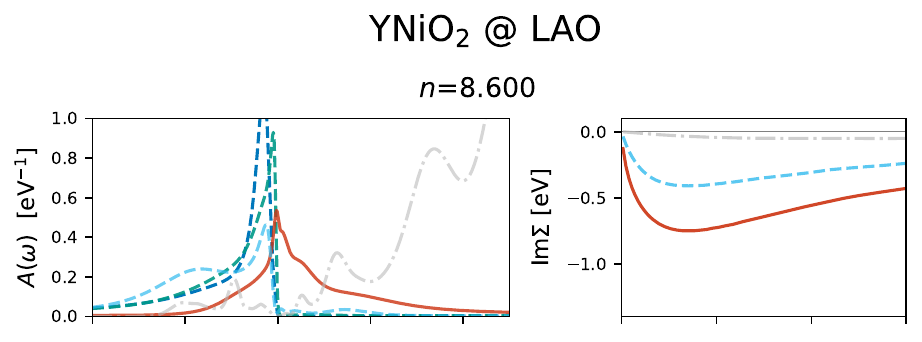}
\includegraphics[angle=0, width=0.54 \columnwidth]{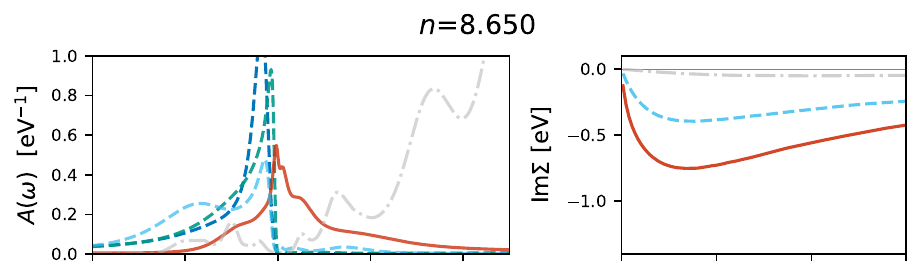}
\includegraphics[angle=0, width=0.54 \columnwidth]{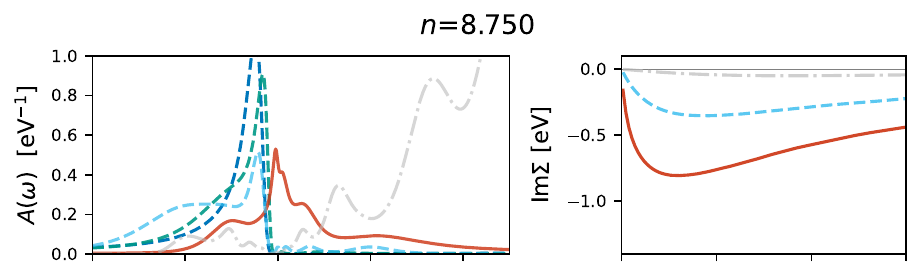}
\includegraphics[angle=0, width=0.54 \columnwidth]{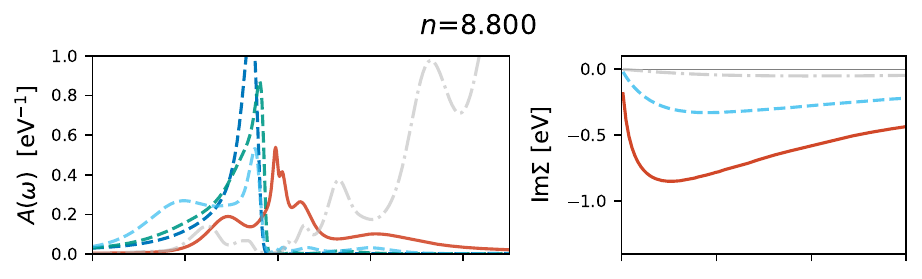}
\includegraphics[angle=0, width=0.54 \columnwidth]{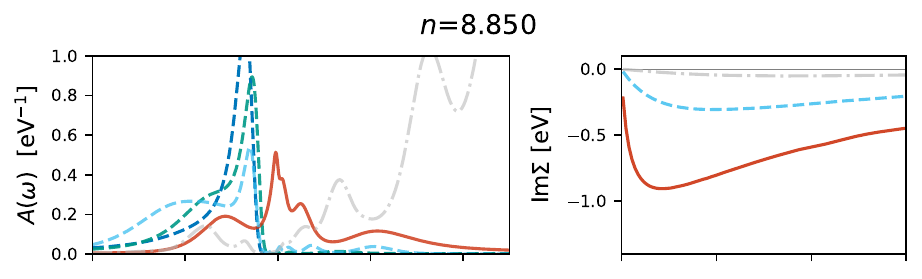}
\includegraphics[angle=0, width=0.54 \columnwidth]{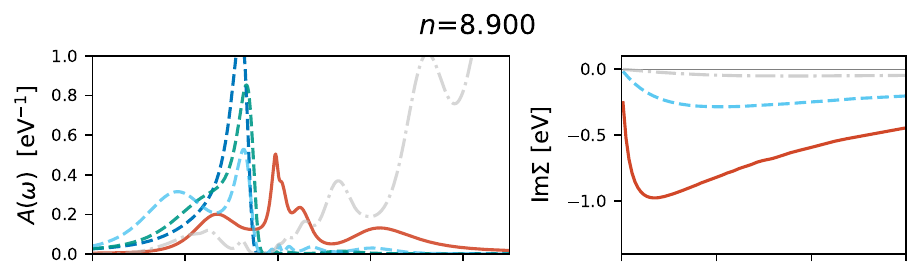}
\includegraphics[angle=0, width=0.54 \columnwidth]{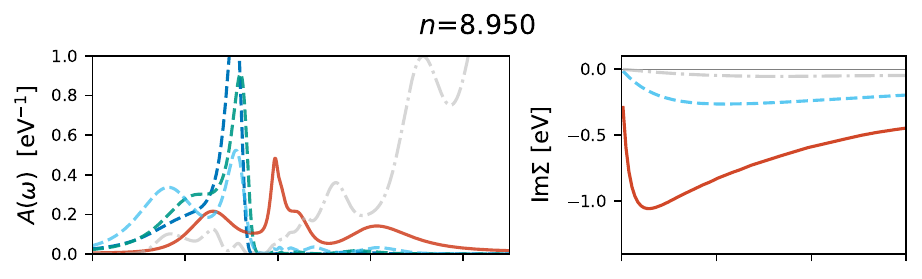}
\includegraphics[angle=0, width=0.55 \columnwidth]{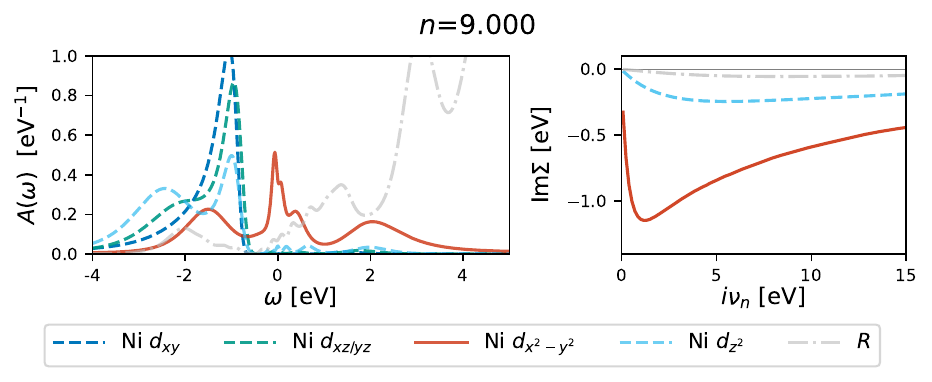}

\caption{\label{Fig:Y_spectra}
Same as Fig.~\ref{Fig:Nd_spectra} for YNiO$_2$@LAO.
}
\end{figure}

\begin{figure}[t]
	\centering
    \vspace{-1cm}
\includegraphics[angle=0, width=0.54 \columnwidth]{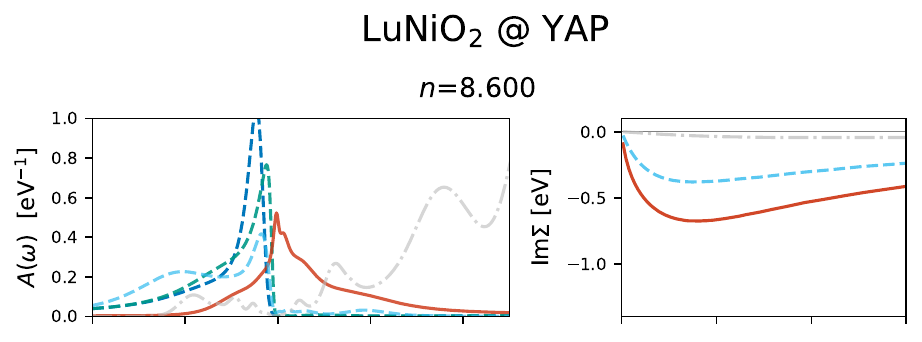}
\includegraphics[angle=0, width=0.54 \columnwidth]{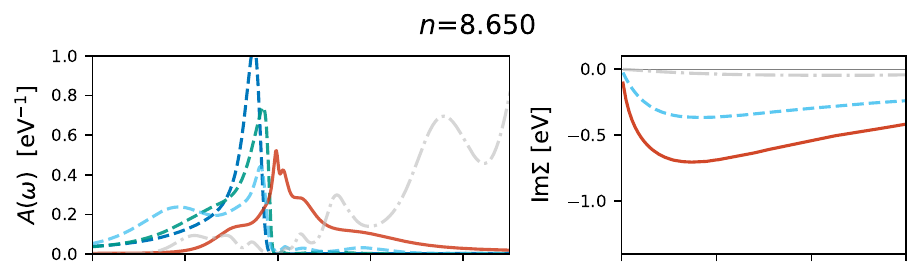}
\includegraphics[angle=0, width=0.54 \columnwidth]{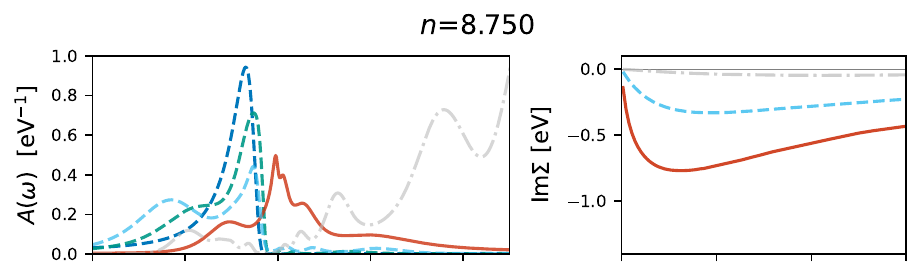}
\includegraphics[angle=0, width=0.54 \columnwidth]{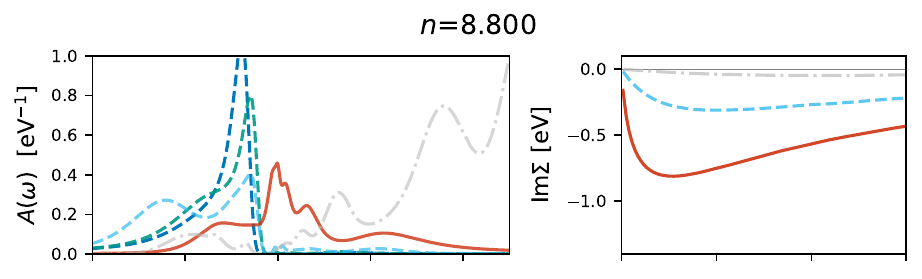}
\includegraphics[angle=0, width=0.54 \columnwidth]{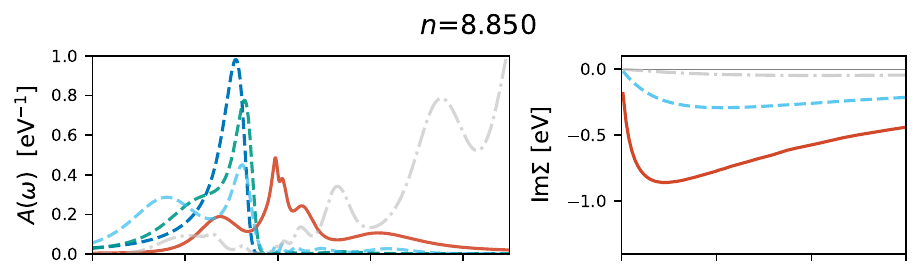}
\includegraphics[angle=0, width=0.54 \columnwidth]{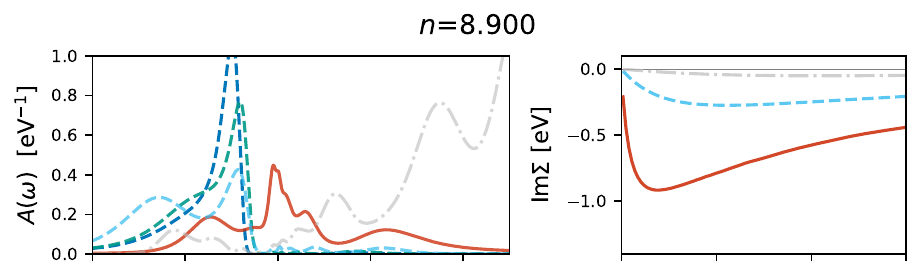}
\includegraphics[angle=0, width=0.54 \columnwidth]{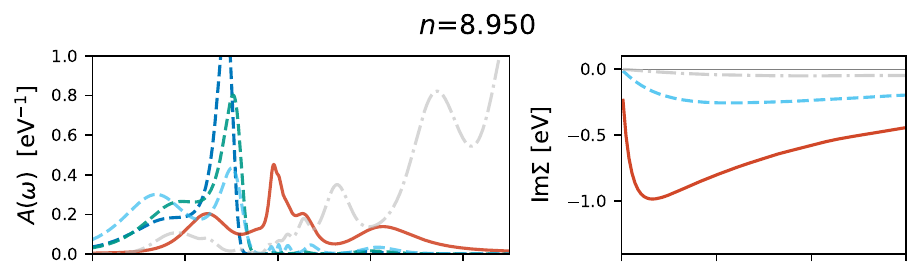}
\includegraphics[angle=0, width=0.55 \columnwidth]{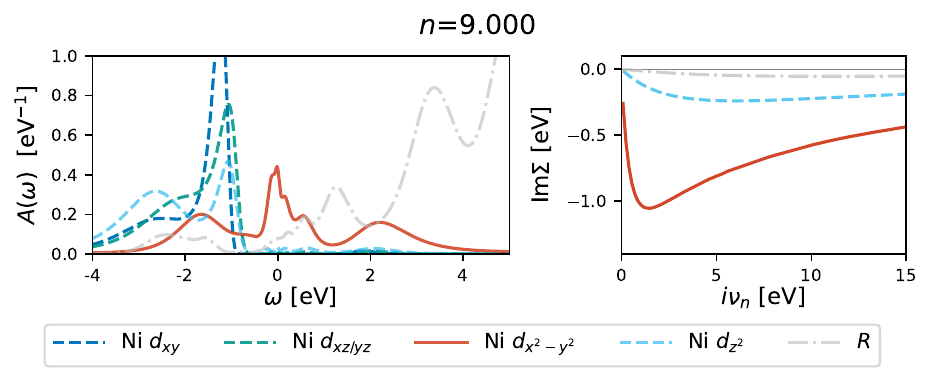}

\caption{\label{Fig:Lu_spectra}
Same as Fig.~\ref{Fig:Nd_spectra} for LuNiO$_2$@YAP.
}
\end{figure}

Figs.~\ref{Fig:Nd_spectra}-\ref{Fig:Lu_spectra} show the (orbital/site-resolved) local DMFT spectral function $A(\omega)$ for the series $R$NiO$_2$ ($R=$Nd,Sm,Y,Lu) considered. The spectra clearly show that at nominal filling (9 electrons) and at low doping, the low-energy physics is dominated by the (close-to-half-filling) $d_{x^2-y^2}$ orbital. The self-doping effect of the $R$ atom is clearly discernible from the $R$ weight (gray dash-dotted line) around $\sim$ -1.5 eV. With an increasing doping $\gtrsim20$\%, this effective single-orbital picture breaks down and the additional Ni $3d$ orbitals become important for the description of the low-energy processes (this picture remains valid for even larger doping levels in the later members of the series).

The corresponding self-energy of the Ni $3d_{z^2}$, Ni $3d_{x^2-y^2}$, and $R$ $5d_{z^2}$ orbitals are also shown as a function of Matsubara frequencies. This avoids the perils of the analytic continuation and is an excellent point of reference for others to compare to or to reproduce our calculations.
Here the (almost empty) $R$ orbitals are --as expected-- very weakly correlated. This is demonstrated by its featureless and almost flat Im$\Sigma(i\nu_n)$. The Ni orbitals instead show the characteristics of a correlated metal in all cases. Going from low to high doping, the almost filled Ni orbitals ($3d_{z^2}$, $3d_{xy}$ and $3d_{xz/yz}$) display an increasingly correlated nature.  At low dopings correlation effects are weak since the orbitals are almost fully filled; correspondingly the self-energy shows the typical behavior of a weakly correlated metal. With larger hole doping, they are activated and the self-energy becomes much larger.

In contrast, the Ni $3d_{x^2-y^2}$ orbital displays the opposite trend, being a strongly correlated metal at the nominal filling, and with increasing doping displaying moderate-to-strong correlations. Interestingly, however, we see the similar effect as discussed in the main text for the single-band model of an effectively less interacting Ni $3d_{x^2-y^2}$ orbital as we go along the Nd$\to$Sm$\to$Y$\to$Lu series at a fixed doping. The imaginary part of the self-energy is less peaked, demonstrating a more weakly (although still moderately-to-strongly) correlated orbital. This is especially clear for the nominal doping, where NdNiO$_2$ has a very sharp renormalized quasiparticle peak at the Fermi energy, which becomes more broadened when going to LuNiO$_2$.
Such a behavior is fully consistent with the expectation of a lower $U/t$ ratio for the effective single-band model, as this encodes an effectively less correlated low-energy model. This result further justifies the use of a fixed $U$ in the $T_c$ calculations in the main text.

\bibliography{main.bib,main2.bib}

\pagebreak

